\documentclass[%
 reprint,
 amsmath,amssymb,
 aps,
 prl,
]{revtex4-2}

\usepackage{graphicx}
\usepackage{float}
\usepackage{dcolumn}
\usepackage{bm}
\usepackage{bbm}
\usepackage{braket}
\usepackage{xcolor}

\newcommand{\commute}[2]{\left[#1,#2\right]}

\usepackage{accents}
\newcommand{\dbtilde}[1]{\accentset{\approx}{#1}}

\newcommand{\ketbra}[2]{\mathinner{|{#1}\rangle\,\langle{#2}|}}

\makeatletter
\DeclareFontFamily{OMX}{MnSymbolE}{}
\DeclareSymbolFont{MnLargeSymbols}{OMX}{MnSymbolE}{m}{n}
\SetSymbolFont{MnLargeSymbols}{bold}{OMX}{MnSymbolE}{b}{n}
\DeclareFontShape{OMX}{MnSymbolE}{m}{n}{
    <-6>  MnSymbolE5
   <6-7>  MnSymbolE6
   <7-8>  MnSymbolE7
   <8-9>  MnSymbolE8
   <9-10> MnSymbolE9
  <10-12> MnSymbolE10
  <12->   MnSymbolE12
}{}
\DeclareFontShape{OMX}{MnSymbolE}{b}{n}{
    <-6>  MnSymbolE-Bold5
   <6-7>  MnSymbolE-Bold6
   <7-8>  MnSymbolE-Bold7
   <8-9>  MnSymbolE-Bold8
   <9-10> MnSymbolE-Bold9
  <10-12> MnSymbolE-Bold10
  <12->   MnSymbolE-Bold12
}{}

\let\llangle\@undefined
\let\rrangle\@undefined
\DeclareMathDelimiter{\llangle}{\mathopen}%
                     {MnLargeSymbols}{'164}{MnLargeSymbols}{'164}
\DeclareMathDelimiter{\rrangle}{\mathclose}%
                     {MnLargeSymbols}{'171}{MnLargeSymbols}{'171}
\makeatother

\begin{document}

\preprint{APS/123-QED}

\title{Non-Markovian transient spectroscopy in cavity QED}

\author{Z.~McIntyre}
\email{zoe.mcintyre@mail.mcgill.ca}
\author{W.~A.~Coish}%
 \email{coish@physics.mcgill.ca}
\affiliation{%
 Department of Physics, McGill University, 3600 rue University, Montreal, QC H3A 2T8, Canada
}%


\date{\today}

\begin{abstract}
We theoretically analyze measurements of the transient field leaving a cavity as a tool for studying non-Markovian dynamics in cavity quantum electrodynamics (QED). Combined with a dynamical decoupling pulse sequence, transient spectroscopy can be used to recover spectral features that may be obscured in the stationary cavity transmission spectrum due to inhomogeneous broadening. The formalism introduced here can be leveraged to perform \emph{in situ} noise spectroscopy, revealing a robust signature of quantum noise arising from non-commuting observables, a purely quantum effect.
\end{abstract}

\maketitle

\emph{Introduction}---Significant effort has recently gone towards reaching the strong-coupling regime of cavity quantum electrodynamics (QED) for individual long-lived spin and charge qubits, with the goals of achieving long-range coupling \cite{vandersypen2017interfacing}, performing fundamental studies of many-body phenomena \cite{desjardins2017observation}, and realizing other exotic effects arising from hybrid systems \cite{clerk2020hybrid}. Strong coupling has been observed between microwave photons and charge qubits in GaAs~\cite{stockklauser2017strong}, spin qubits in silicon \cite{mi2017strong, mi2018coherent, samkharadze2018strong}, resonant-exchange qubits in GaAs triple quantum dots~\cite{landig2018coherent}, and spin qubits in carbon nanotube double quantum dots~\cite{cubaynes2019highly, viennot2015coherent} (DQDs). Two-qubit, photon-mediated interactions have been observed between charge qubits in GaAs DQDs~\cite{vanwoerkom2018microwave} and between spin qubits in silicon DQDs~\cite{borjans2020resonant, harveycollard2021circuit}. As these devices reach a progressively higher level of sophistication and quality, it is increasingly important to characterize the qubits and their local environments \emph{in situ}, together with the components that define the cavity.

\begin{figure}
    \centering
    \includegraphics[width=\linewidth]{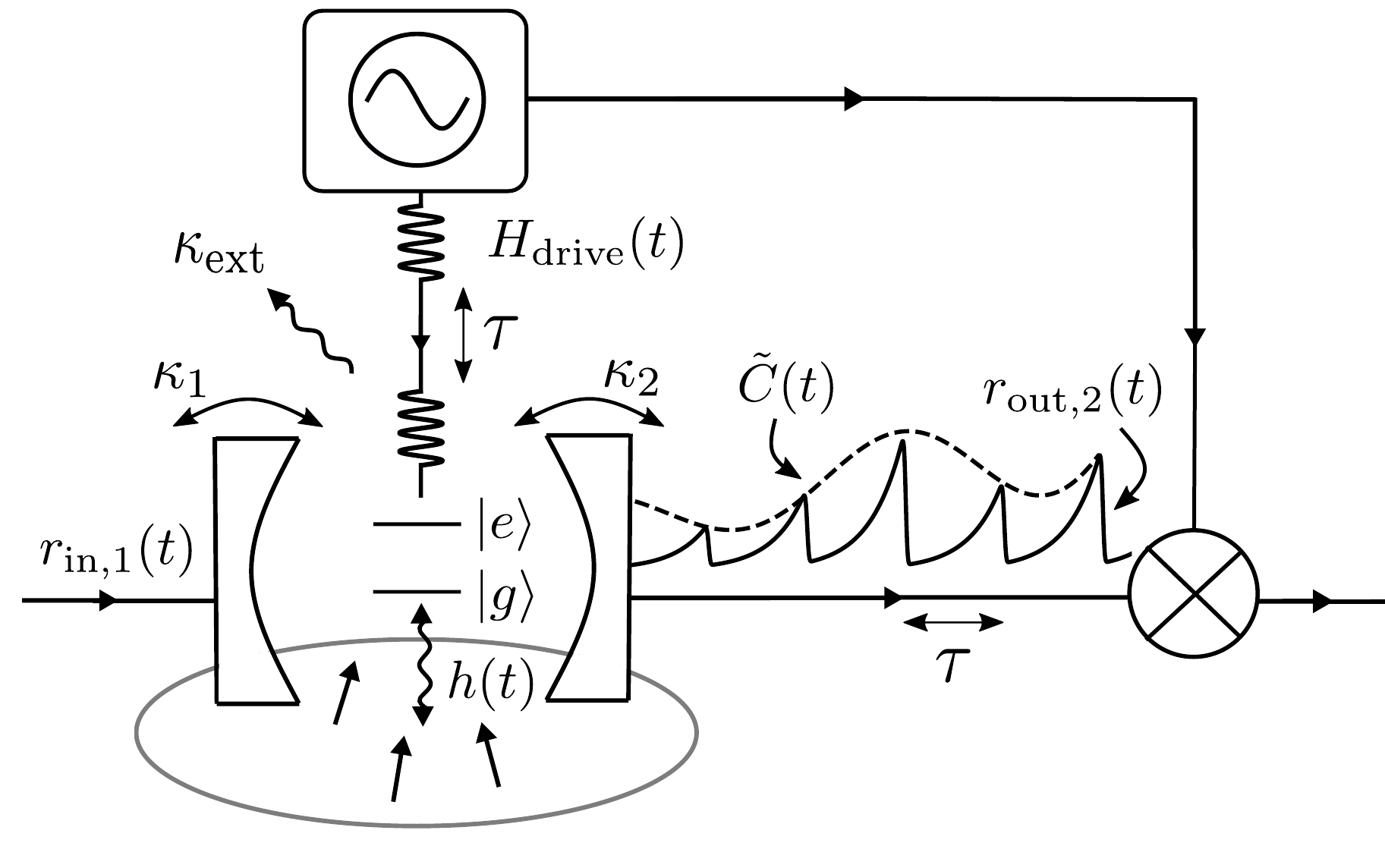}
    \caption{A typical cavity-QED setup. The transmission spectrum is obtained from the linear response of the output signal $r_\mathrm{out,2}(t)$ to a monochromatic input tone, $r_{\mathrm{in,1}}(t)$. In contrast, the transient spectrum is found, for $r_{\mathrm{in,1}}(t)=0$, by generating a sequence of control pulses via a qubit drive $H_{\mathrm{drive}}(t)$. These pulses induce non-Markovian coherence revivals with envelope $\tilde{C}(t)$ for a qubit coupled to its environment through some interaction $h\sigma_z/2$. A phase-sensitive measurement of $r_{\mathrm{out},2}(t)$ can then be used to determine $\tilde{C}(t)$. The decay rates at the input and output ports of the cavity are denoted $\kappa_1$ and $\kappa_2$, respectively, while extrinsic decay is denoted $\kappa_{\mathrm{ext}}$.}
    \label{fig:cavity-schematic}
\end{figure}

\emph{In-situ} characterization of a two-level emitter (qubit) coupled to a cavity is often done by measuring a transmission or reflection spectrum \cite{hood1998real,ye1999trapping,stockklauser2017strong,mi2017strong,samkharadze2018strong,blais2021circuit} in a setup similar to that shown in Fig.~\ref{fig:cavity-schematic}. In this setup, an input tone $r_{\mathrm{in},1}(t)=(2\pi)^{-1}\int d\omega e^{-i\omega t}r_{\mathrm{in},1}(\omega)$ is introduced, and after a time long compared to the cavity decay time $\kappa^{-1}$, the output field $r_{\mathrm{out},2}(t)$ reaches a steady state. The stationary transmission $A_\mathrm{T}(\omega)=r_{\mathrm{out},2}(\omega)/r_{\mathrm{in},1}(\omega)$ then carries information about the qubit accounting for its interaction with the environment and resulting decay processes. To interpret the transmission, it is common to make the simplifying assumption that the qubit dynamics are generated by a Markovian master equation, with parameters characterizing dephasing and relaxation rates. The standard tools of input-output theory \cite{gardiner1985input, jacobs2014quantum,burkard2020superconductor,blais2021circuit} can then be applied. The Markovian assumption is often an excellent approximation for single-atom emitters \cite{hood1998real,ye1999trapping} and for the superconducting transmon qubits commonly used in circuit QED~\cite{blais2021circuit}. With some exceptions \cite{schlor2019correlating,burnett2019decoherence}, these systems typically have coherence times $T_2$ limited by the exponential energy relaxation time $T_1$: $T_2\simeq 2T_1$. In stark contrast, spin and charge qubits defined using semiconductor nanostructures almost universally undergo non-exponential (non-Markovian) pure dephasing on a time scale $T_2^*\ll T_1$ arising from inhomogeneous broadening due to low-frequency charge noise or slow nuclear-spin environments. A different approach is required for these and many other non-Markovian systems. 

A dynamical decoupling pulse sequence can help mitigate the effects of strong inhomogeneous broadening, but the result is a train of manifestly non-Markovian collapses and revivals (echoes) in qubit coherence. Although these revivals (of duration $\sim T_2^*$) can be mapped to the transient output field $r_{\mathrm{out},2}(t)$ (see Fig.~\ref{fig:cavity-schematic}), their effect on the stationary transmission spectrum is negligible since they are, by definition, transient. Similar revivals have already been exploited for measurement in spin-echo experiments on ensembles \cite{ranjan2020pulsed,ledantec2021twenty}. For a low-$Q$ resonator, the relationship between spin coherence and the output field is relatively simple and time-local \cite{ranjan2020pulsed}. By contrast, it is a nontrivial problem to relate the complex pattern of revivals arising from, e.g., a dynamical decoupling sequence, to real-time non-Markovian coherence dynamics for a high-$Q$ cavity. We perform this analysis here. Although our focus is on individual spin and charge qubits under a widely used dynamical decoupling sequence, the ideas presented here are generally applicable to ensembles and to a wide range of non-Markovian systems in cavity QED, a topic of significant recent interest~\cite{vcernotik2019cavity, bundgaard2021non, carmele2019non, sinha2020non, krimer2014nonMarkovian}.  

\emph{Model}---We start from a typical cavity-QED setup  (see Fig.~\ref{fig:cavity-schematic}), with dynamics governed by the quantum master equation (taking $\hbar=1$): 
\begin{equation}
    \dot{\rho}=-i[H(t),\rho]+\frac{\gamma_\phi}{2}\mathcal{D}[\sigma_z]\rho+\kappa_{\mathrm{ext}}\mathcal{D}[a]\rho.\label{eq:master-eq}
\end{equation}
Here, $\rho=\rho(t)$ is the joint state of the qubit, cavity, quantum environment, and transmission lines. The qubit (with Pauli operator $\sigma_z=\ket{e}\bra{e}-\ket{g}\bra{g}$) undergoes Markovian pure dephasing at a rate $\gamma_\phi$, while photons in the cavity mode (annihilated by $a$) decay at an extrinsic rate $\kappa_\mathrm{ext}$. The damping superoperator acts like $\mathcal{D}[\mathcal{O}]\rho=\mathcal{O}\rho\mathcal{O}^\dagger-\{\mathcal{O}^\dagger\mathcal{O},\rho\}/2$ for any operator $\mathcal{O}$; in addition to damping processes, the density operator $\rho$ evolves under the Hamiltonian
\begin{align}
H(t)&=\frac{1}{2}[\Delta+\Omega(t)]\sigma_z+\frac{1}{2}h(t)+\omega_c a^\dagger a+g\sigma_x(a+a^\dagger)\nonumber\\
&+H_{\mathrm{drive}}(t)+\sum_{i=1,2}\sum_k(\lambda_{k,i}e^{i\omega_k t}r_{k,i}^\dagger a+\mathrm{h.c.}),\label{eq:hamiltonian}
\end{align}
where $\Delta$ is the qubit resonance frequency, $\omega_c$ is the cavity frequency, and $g$ is the qubit-cavity coupling. The term $H_{\mathrm{drive}}(t)$ describes a drive acting on the qubit, while noise is generated by $\Omega(t)=\eta(t)+h(t)$, where $\eta(t)$ is a classical noise parameter and $h(t)=e^{i(H_{\mathrm{E}}-h/2)t}he^{-i(H_{\mathrm{E}}-h/2)t}$ acts on the environment alone. The time dependence of $h(t)$ arises from a lab-frame Hamiltonian $H_{\mathrm{E}}+h\sigma_z/2$ together with the assumption that the environment is prepared in a steady state while coupled to the qubit held in $\ket{g}$ (see below and Ref.~\onlinecite{supplement}). We take $\eta(t)$ to be generated by a stationary Gaussian process with zero mean ($\llangle\eta(t)\rrangle=0$) and spectral density 
\begin{equation}
    S_\eta(\omega)=\int dt e^{-i\omega t}\llangle\eta(t)\eta(0)\rrangle,\label{eq:classical-noise-spectral-density}
\end{equation}
where here, the double angle brackets $\llangle\rrangle$ represent an average over noise realizations. The terms $\propto\lambda_{k,i}$ in Eq.~\eqref{eq:hamiltonian} describe coupling of the cavity mode to the input (output) [for $i=1(2)$] transmission-line mode annihilated by $r_{k,i}$ and having frequency $\omega_k$. For modes propagating in one dimension, $r_{\mathrm{in},i}(t)=[c/L]^{1/2}\sum_ke^{-i\omega_kt}\braket{r_{k,i}}_0$ and $r_{\mathrm{out},i}(t)=[c/L]^{1/2}\sum_k\braket{r_{k,i}}_t$, where $L$ is the length of the transmission line and $c$ is the speed of light. The notation $\braket{\mathcal{O}}_t$ indicates an average with respect to the state $\rho(t)$, together with an average over realizations of the classical noise $\eta(t)$: $\braket{\mathcal{O}}_t=\llangle\mathrm{Tr}\{\mathcal{O}\rho(t)\}\rrangle$.

\emph{Transient spectroscopy}---In order to accurately monitor qubit dynamics through the transient output field $r_{\mathrm{out},2}(t)$, we consider the following protocol: (i) An undriven single-sided ($\kappa_1=0$) cavity is prepared in a vacuum state $\ket{0}$ far-detuned from (or decoupled from) the qubit. (ii) The qubit is prepared in its ground state $\ket{g}$, and the environment is allowed to reach a steady-state $\bar{\rho}_\mathrm{E}$ in contact with the qubit: $[H_{\mathrm{E}}-h/2,\bar{\rho}_{\mathrm{E}}]=0$. (iii) At $t=0$, the qubit and cavity are tuned close to resonance (or the coupling $g$ is turned on), and a finite drive $H_\mathrm{drive}(t\ge 0)$ generates qubit coherence $\braket{\sigma_x}_t$. This coherence is related to the cavity field $\braket{\tilde{a}}_t=e^{i\Delta t}\braket{a}_t$ via direct integration of Eq.~\eqref{eq:master-eq}:
\begin{equation}
    \braket{\tilde{a}}_t=-ig\int_{-\infty}^{\infty} dt'\:\chi_c(t-t')e^{i\Delta t'}\braket{\sigma_x}_{t'},\quad\label{eq:cavity-output}
\end{equation}
where $\chi_c(t)=e^{-i\delta t-\kappa t/2}\Theta(t)$ for a cavity-qubit detuning $\delta=\omega_c-\Delta$ and total cavity decay rate $\kappa=\kappa_{\mathrm{ext}}+\kappa_2$. Neglecting retardation effects, the measured output field is then given by the input-output relation $r_{\mathrm{out},2}(t)=\sqrt{\kappa_2}\braket{a}_t$~\cite{gardiner1985input}.  For a single cavity-coupled qubit, the protocol [(i)-(iii)] is limited to gathering $\lesssim 1$ bit of information per cycle, similar to many early cavity-QED schemes \cite{hood1998real}. These steps must therefore be repeated many times to estimate the expectation value $\braket{\sigma_x}_t$.

\emph{Dynamical decoupling}---For concreteness, we consider an $N$-pulse Carr-Purcell-Meiboom-Gill (CPMG) sequence, where coherence preparation at $t=0$ [such that $\braket{\sigma_-}_0=\frac{1}{2}\braket{\sigma_x}_0\ne 0$] is followed by $\pi_x$-pulses at times $t=\tau/2, 3\tau/2,...,(N-1/2)\tau$, leading to coherence revivals at times $t=n\tau$, $n=1,2,\dots,N$ (see the supplement~\cite{supplement} for the general formalism, valid for other pulse sequences). We further specialize to the regime $g<\kappa\ll\tau^{-1}$, where cavity backaction effects can be treated as a small correction. In this regime, the coherence factor $C(t)=\braket{\sigma_-}_t/\braket{\sigma_-}_0$ can be written in terms of a comb of revivals (echoes) with peaks $\sim G_n(t-n\tau)$ centered at $t=n\tau$ and an echo envelope $\tilde{C}(t)$: 
\begin{equation}
C(t)=\sum_ne^{-i\Delta(t-n\tau)}G_n(t-n\tau)\mathcal{K}^n\tilde{C}(n\tau).\label{eq:coherence-factor}
\end{equation}
Here, $\mathcal{K}z=z^*$ for all $z\in\mathbb{C}$. If $T_2^*\ll\tau$ [where $2/(T_2^*)^2=(2\pi)^{-1}\int d\omega\:S_\eta(\omega)$], and if $\tilde{C}(n\tau)$ is slowly varying on the timescale $T_2^*$, then we find
\begin{equation}
   G_n(t)=e^{\sqrt{\gamma_\mathrm{P}n\tau}}\left\llangle e^{-\Gamma_\mathrm{P}(\eta)n\tau/2}e^{-i\eta t}\right\rrangle, 
\end{equation}
where $\eta=\eta(0)$ is the low-frequency contribution to $\eta(t)$,  $\Gamma_\mathrm{P}(\eta)=g^2\kappa/[(\eta-\delta)^2+(\kappa/2)^2]$ is the Purcell decay rate at fixed $\eta$, and $\gamma_\mathrm{P}=(gT_2^*)^2\kappa/2$. The echo envelope is
\begin{equation}
    \tilde{C}(n\tau)=e^{-\sqrt{\gamma_\mathrm{P}n\tau}-\gamma_\phi n\tau}\llangle\mathrm{Tr}\{U_-^\dagger(n\tau)U_+(n\tau)\bar{\rho}_E\}\rrangle,
\end{equation}
where $U_{\pm}(n\tau)=\mathcal{T}\mathrm{exp}\{-\frac{i}{2}\int_0^{n\tau} [h(t')\pm s(t')\Omega(t')]\}$ are evolution operators acting on the environment, conditioned on the $\sigma_z$-eigenvalue $(\pm)$ of the qubit. Here, $\mathcal{T}$ is the time-ordering operator and $s(t)=(-1)^{n(t)}$ for $n(t)$ $\pi$-pulses having taken place up to time $t$. For $n<1/(\gamma_\mathrm{P}\tau)$, backaction due to Purcell decay is negligible and the revivals are well approximated by $G_n(t)\simeq G_0(t)=e^{-\left(t/T_2^*\right)^2}$. The effects of backaction will be further discussed below.

Taking the Fourier transform of Eq.~\eqref{eq:cavity-output} gives $\braket{\tilde{a}}_\omega=-ig\chi_c(\omega)\braket{\sigma_x}_{\omega+\Delta}$; the cavity susceptibility acts as a filter, $\chi_c(\omega)=[i(\delta-\omega)+\kappa/2]^{-1}$. In the high-$Q$ limit ($Q=\omega_c/\kappa\gg 1$), $\chi_c(\omega)$ suppresses the counter-rotating component $\braket{\sigma_+}_t$, allowing us to replace $\braket{\sigma_x}_t\simeq \braket{\sigma_-}_t=C(t)\braket{\sigma_-}_0$ in Eq.~\eqref{eq:cavity-output} for $|\delta|\ll |\Delta|$. Under the assumptions laid out above, we find a general expression relating $\braket{\tilde{a}}_\omega$ to the echo envelope $\tilde{C}(n\tau)$ \cite{supplement}. For a narrow cavity resonance, $\kappa T_2^*\ll 1$, $\braket{\tilde{a}}_\omega$ will be sharply peaked around $\omega=\delta$, leading to:
\begin{equation}
    \braket{\tilde{a}}_{\omega=\delta}\simeq-i\braket{\sigma_x}_0\frac{\sqrt{\pi}gT_2^*}{\kappa}\Big[\tilde{C}_{N,\tau}(\delta)-\frac{1}{2}C(0)\Big],\label{eq:spectrum-peak}
\end{equation}
where we neglect corrections smaller by $O(g/\kappa)$, $O(\kappa/\Delta)$. Here,
\begin{equation}
   \tilde{C}_{N,\tau}(\omega)=\sum_{n=0}^{N}e^{in(\omega+\Delta)\tau}\bar{G}_n\mathcal{K}^n\tilde{C}(n\tau), 
\end{equation}
where $\bar{G}_n=(\sqrt{\pi} T_2^*)^{-1}\int_{-\infty}^\infty dt G_n(t)$. The echo envelope $\tilde{C}(t)$ can thus be reconstructed by measuring $r_{\mathrm{out},2}(\omega)$ to infer $\braket{\tilde{a}}_{\omega=\delta}$ over some detuning interval $O(2\pi/\tau)$ and then inverting the discrete Fourier transform $\tilde{C}_{N,\tau}(\omega)$. The revival amplitudes $\mathcal{K}^n\tilde{C}(n\tau)$ depend alternately on $\tilde{C}(n\tau)$ (for $n$ even) and $\tilde{C}^*(n\tau)$ (for $n$ odd); this even/odd alternation is a direct result of the high-$Q$ limit, which (as we now show) can be exploited to identify a purely quantum effect.

\emph{Quantum noise}---For the secular coupling $h\sigma_z/2$ considered here, we find a generic expression for the echo envelope $\tilde{C}(n\tau)$:  Without loss of generality, we take $\braket{h}_t=0$, in which case a Magnus expansion to second order in $h(t)$ followed by a Gaussian approximation (valid for a large uncorrelated environment \cite{beaudoin2013enhanced}---see Ref.~\onlinecite{norris2016qubit} for the non-Gaussian generalization) gives
\begin{equation}
    \tilde{C}(n\tau)\simeq e^{-\sqrt{\gamma_{\mathrm{P}}n\tau}-\gamma_\phi n\tau} \mathrm{exp}\{-i\Phi_{\mathrm{q}}(n\tau)-\chi(n\tau)\},
\end{equation}
where
\begin{align}
&\Phi_{\mathrm{q}}(n\tau)=\int \frac{d\omega}{2\pi}\frac{F_{\mathrm{q}}(\omega,n\tau)}{\omega^2}S_{\mathrm{q}}(\omega),\label{eq:phi}\\
&\chi(n\tau)=\int \frac{d\omega}{2\pi}\frac{F_{\mathrm{c}}(\omega,n\tau)}{\omega^2}S_{\mathrm{c}}(\omega).\label{eq:chi}
\end{align}
Here, $F_{\mathrm{c}}(\omega,n\tau)=(\omega^2/2)\left|\int_0^{n\tau} dt \:e^{i\omega t}s(t)\right|^2$ is the usual filter function for classical noise \cite{cywinski2008how}, $F_{\mathrm{q}}(\omega,n\tau)=\omega\int_0^{n\tau}dt\:\mathrm{sin}(\omega t)s(t)$ is a new quantum-noise filter function, and 
\begin{align}
    S(\omega)&=S_{\mathrm{c}}(\omega)+iS_{\mathrm{q}}(\omega)\nonumber\\
    &=\lim\limits_{\epsilon\rightarrow 0^+}\int_{-\infty}^\infty dt\:e^{-i\omega t-\epsilon\lvert t\rvert}\braket{\Omega(\lvert t\rvert)\Omega}\label{eq:spectral-density}
\end{align}
is the spectral density~\cite{supplement}.
The magnitude of $\tilde{C}(n\tau)$ is then determined by the classical part of the noise spectrum $S_\mathrm{c}(\omega)=\mathrm{Re}\left[S(\omega)\right]=S_\eta(\omega)+S_h(\omega)$, where $S_h(\omega)$ depends only on the symmetrized correlation function $\left<\left\{h(\lvert t\rvert),h\right\}\right>$. For a quantum environment, the envelope $\tilde{C}(n\tau)$ generally has a phase $\Phi_{\mathrm{q}}(n\tau)$ [Eq.~\eqref{eq:phi}] determined by the quantum noise $S_\mathrm{q}(\omega)=\mathrm{Im}\left[S(\omega)\right]$, due to the antisymmetrized correlation function $\left<\left[h(\lvert t\rvert),h\right]\right>$ \cite{supplement}. This phase will manifest itself in the alternation between $\tilde{C}$ and $\tilde{C}^*$ in the discrete Fourier transform in Eq.~\eqref{eq:spectrum-peak} for $n$ even/odd. The importance of quantum noise due to non-commuting observables has long been recognized in the mesoscopic-physics community \cite{gavish2000detection}. In addition, it has been measured in CPMG experiments performed on nitrogen-vacancy center spin qubits in diamond, leading to a phase shift $\Phi_{\mathrm{q}}(n\tau)\sim\pi$ \cite{zhao2012sensing}. 

Despite this recognition of quantum noise in other communities, a common simplification in noise spectroscopy is to assume a frequency-symmetric, real-valued spectrum, $S(\omega)=S(-\omega)=S^*(\omega)$, as would arise from a classical fluctuating field \cite{alvarez2011measuring,yuge2011measurement,bylander2011noise,szankowski2017environmental, szankowski2019transition, chan2018assessment} (although quantum noise has been incorporated into some recent theory works \cite{pazsilva2017multiqubit,kwiatkowski2020influence}). By leveraging the sensitivity of the cavity field to the phase of qubit coherence revivals in the high-$Q$ regime [Eq.~\eqref{eq:spectrum-peak}], we identify a robust even-odd modulation of revivals arising from $S_q(\omega)=\mathrm{Im}\:S(\omega)$, unique to quantum environments. Notably, this even-odd effect would not appear for coupling of the form $h\sigma_z/2$ when $\bar{\rho}_{\mathrm{E}}$ is stationary with respect to $H_{\mathrm{E}}$ alone ($[H_{\mathrm{E}},\bar{\rho}_{\mathrm{E}}]=0$) ~\cite{kwiatkowski2020influence, pazsilva2017multiqubit}, as may occur for an environment prepared in the absence of the qubit.  The quantum-noise phase $\Phi_{\mathrm{q}}(n\tau)$ [Eq.~\eqref{eq:phi}] thus appears as a direct consequence of the initial condition, $[H_{\mathrm{E}}-h/2,\bar{\rho}_{\mathrm{E}}]=0$~\cite{supplement}. 

\begin{figure}
    \centering
    \includegraphics[width=\linewidth]{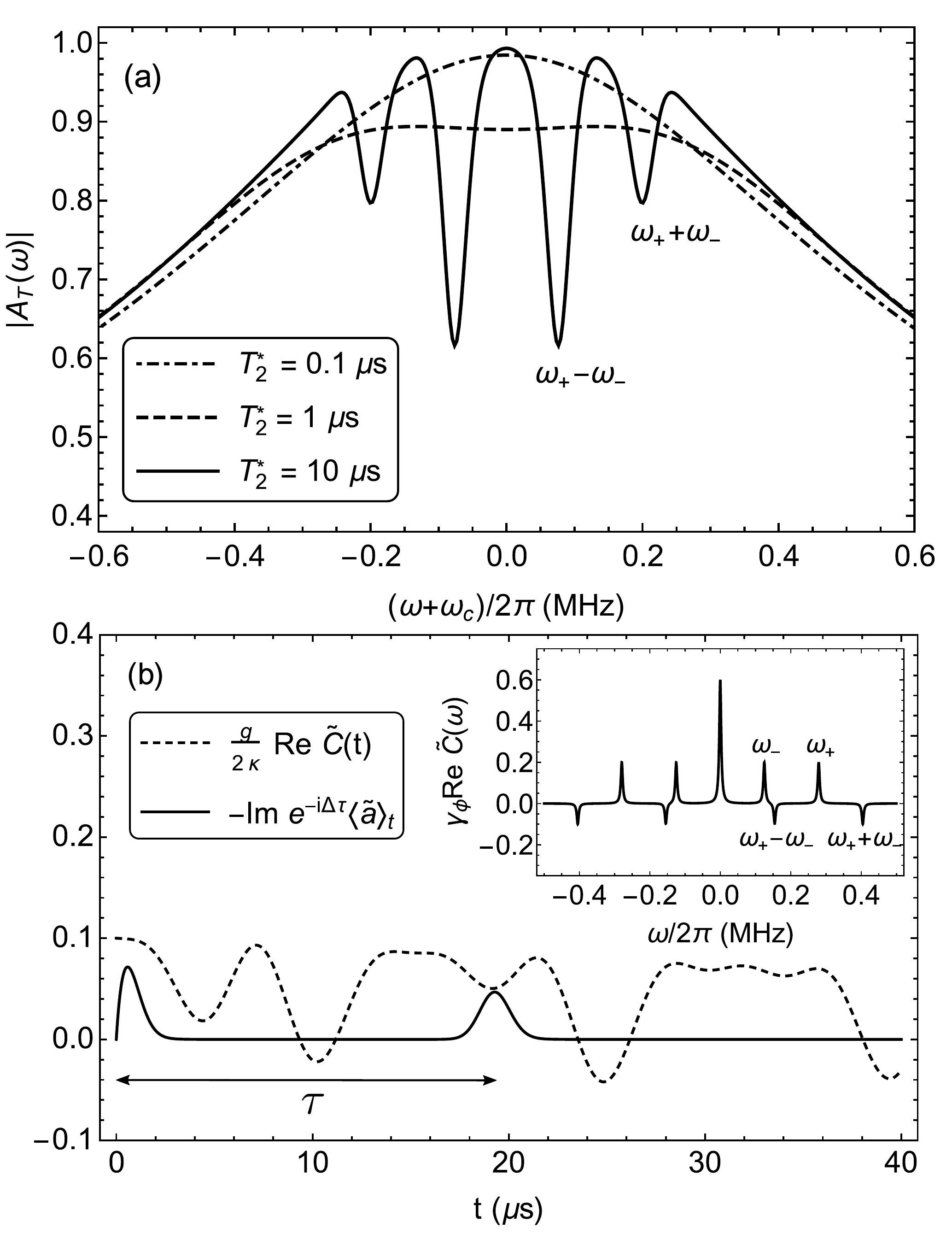}
    \caption{(a) Inhomogeneously broadened cavity transmission at $\delta=0$ for three values of $T_2^*$ ($T_2^*=0.1$ $\mu$s, 1 $\mu$s, and 10 $\mu$s). We take $A/2\pi=-0.250\:\mathrm{MHz}$~\cite{hensen2020silicon}, $\gamma B_z=\gamma B_x=A/2$, $\gamma_\phi^{-1}=100\;\mu\mathrm{s}$, $\kappa/2\pi=1\:\mathrm{MHz}$, and $g/\kappa=0.2$. For $B_z=15$ mT,  $\omega_c/2\pi=g^*\mu_\mathrm{B}B_z/2\pi=0.4$ GHz, where $g^*=2$. We assume an infinite-temperature initial state for the nuclear spin. (b) Revivals in the cavity field, modulated by the echo envelope, assuming the same parameters as in (a) with $T_2^*=1\:\mu\mathrm{s}$. Once the echo envelope has been mapped out (e.g.~by varying $\tau$), it can be Fourier transformed [inset] to recover the frequencies $\omega_\pm$ obscured in (a).} \label{fig:inhomogeneous-broadening}
\end{figure}

\emph{Characterizing a single nuclear spin}---As a concrete application, we consider an electron spin qubit in a silicon double quantum dot (DQD)~\cite{mi2017strong, mi2018coherent, samkharadze2018strong}, exposed to a spatially varying magnetic field. The magnetic field is assumed to have an $x$-component $B_x(\boldsymbol{r})$ that averages to zero over the DQD and a uniform $z$-component $B_z$~\cite{beaudoin2013enhanced}, a setup that is commonly used to generate spin-photon coupling \cite{beaudoin2016coupling}. This leads to a secular coupling and ``environment" Hamiltonian given by
\begin{equation}
    \frac{1}{2}h\sigma_z= \frac{1}{2}A I_z\sigma_z,\quad H_{\mathrm{E}}=\gamma\left(B_xI_x+B_zI_z\right).\label{eq:hamiltonian-hyperfine}
\end{equation}
Here, $B_x=B_x(\boldsymbol{r}_0)$ for a $^{29}$Si nuclear spin located at position $\boldsymbol{r}_0$, $\bm{I}$ is a spin-I operator ($I=1/2$ for ${}^{29}$Si), $A$ is the hyperfine coupling, and $\gamma=-5.319\times 10^7\,\mathrm{rad}\,\mathrm{T}^{-1}\,\mathrm{s}^{-1}$ is the gyromagnetic ratio. The same model also applies to a spin qubit in a uniform $\mathbf{B}$-field, provided the spin has an anisotropic $g$-tensor, leading to non-collinear quantization axes for the qubit and nuclear spin. Alternatively, this model can describe a qubit having a finite charge dipole interacting with the electric field produced by a two-level charge fluctuator, where $\gamma B_z$ and $\gamma B_x$ are replaced by the fluctuator bias and tunnel splitting, respectively \cite{shnirman2005low,galperin2006nonGaussian,schlor2019correlating}. 

For an electron-spin qubit in isotopically enriched silicon, coherence times may be limited by a small number of ${}^{29}$Si nuclear spins \cite{zhao2019single}. Extracting parameters for individual nuclear spins could facilitate decoherence suppression through a notch-filter dynamical decoupling sequence \cite{malinowski2017notch}, or allow for a transfer of information from the electron spin to the nuclear spin for a long-lived quantum memory. The problem of characterizing the spin state of a single ${}^{31}\mathrm{P}$ donor nuclear spin (with hyperfine coupling $A/2\pi\approx 25$ MHz) was recently considered theoretically in Ref.~\onlinecite{mielke2021nuclear} in the context of transmission spectroscopy (described by input-output theory). For a ${}^{29}\mathrm{Si}$ nuclear spin coupled to a quantum-dot-bound electron spin, however, the hyperfine coupling is orders of magnitude weaker ($A/2\pi\approx-0.25$ MHz has been measured, for instance \cite{hensen2020silicon}). Spectral information in the transmission $A_\mathrm{T}(\omega)=\llangle A_\mathrm{T}(\omega,\eta)\rrangle$ might be entirely obscured due to inhomogeneous broadening as a result [Fig.~\ref{fig:inhomogeneous-broadening}(a)]. Here, we have averaged the transmission $A_\mathrm{T}(\omega,\eta)$ for a time-independent but random value of $\eta$ \cite{supplement}. Even when spectral information about the nuclear spin is completely obscured in a more conventional measurement of the transmission spectrum, it can still be recovered in the transient spectrum resulting from a spin-echo sequence.
 
A finite value $B_x\ne 0$ leads to echo envelope modulations following a Hahn-echo sequence (CPMG with $N=1$). For $\kappa T_2^*> 1$ but $Q=\omega_c/\kappa\gg 1$, the cavity-field revival is modulated by the echo envelope according to $\braket{a}_{\tau}\simeq -i\frac{2g}{\kappa}\tilde{C}(\tau)\braket{\sigma_-}_0$ [see Fig.~\ref{fig:inhomogeneous-broadening}(b)]. [In the opposite regime, $\kappa T_2^*< 1$, it is instead modulated according to $\braket{a}_\tau\simeq -i\sqrt{\pi}gT_2^*\tilde{C}(\tau)\braket{\sigma_-}_0$.] In the case of a single environmental spin, the Gaussian approximation cannot be justified, but this model can be solved exactly: For a fully randomized (infinite-temperature) initial condition for the nuclear spin, the Hahn-echo amplitude at $t=\tau$ is given by $\tilde{C}(\tau)=e^{-\gamma_\phi\tau}[1-\delta \tilde{C}(\tau)]$, where
\begin{equation}
    \delta \tilde{C}(\tau)=2\sin^2(\Delta\phi)\sin^2\left(\frac{\omega_+\tau}{4}\right)\sin^2\left(\frac{\omega_-\tau}{4}\right).\label{eq:eseem-envelope}
\end{equation}
Here, $\omega_\pm=[(\gamma B_x)^2+(\gamma B_z\pm A/2)^2]^{1/2}/2$ and $\Delta\phi=\phi_+-\phi_-$, where $\phi_{\pm}=\mathrm{arctan}[2\gamma B_x/(2\gamma B_z\pm A)]$. For the infinite-temperature environmental initial condition considered here, $\tilde{C}(\tau)$ is real and there is no quantum-noise contribution. A polarized initial condition would, however, lead to a complex-valued $\tilde{C}(\tau)$, a signature of a quantum environment and of quantum noise~\footnote{For $h(t)=e^{i(H_{\mathrm{E}}-h/2)t}he^{-i(H_{\mathrm{E}}-h/2)t}=\sum_\alpha c_\alpha(t) I_\alpha$, the quantum-noise term produces a phase arising from $\braket{[h(t),h]}=i\sum_{\alpha\beta}\epsilon_{\alpha\beta\gamma}c_\alpha(t)c_\beta(0)\braket{I_\gamma}$, which is nonzero for a polarized initial state, $\braket{I_\gamma}=\mathrm{Tr}\left(\bar{\rho}_\mathrm{E}I_\gamma\right)\neq 0$. This analysis applies within the Gaussian approximation, which can be justified even for a single spin at short times.}. In this illustrative case of coupling to a single spin, the frequencies $\omega_\pm$ and angular difference $\Delta\phi$ can be extracted independently from the peak positions and peak heights in a Fourier transform of Eq$.$~(\ref{eq:eseem-envelope}) [Fig$.$~\ref{fig:inhomogeneous-broadening}(b), inset], allowing for recovery of both components of the local magnetic field $B_x,B_z$, and of the local hyperfine coupling $A$. For the value of $A$ used in Fig.~\ref{fig:inhomogeneous-broadening}, the visibility $\mathrm{sin}^2(\Delta\phi)$ of the echo-envelope oscillations [c.f.~Eq.~\eqref{eq:eseem-envelope}] is maximized for $B_z=B_x=15$ mT. While this combination of values is possible, it would be fortuitous and would require a relatively low cavity frequency, $\omega_c/2\pi\simeq 0.4\,\mathrm{GHz}$. Away from these values, $\mathrm{sin}^2(\Delta\phi)\simeq [A\gamma B_x/(\gamma B_z)^2]^2$ for $A,\gamma B_x<\gamma B_z$. When $[A\gamma B_x/(\gamma B_z)^2]^2\ll 1$, the amplitude of the modulations can be enhanced through a large-$N$ CPMG sequence: As is well known, a multi-pulse CPMG sequence can be used to amplify specific Fourier components of the noise \cite{taylor2008high,szankowski2017environmental,zhao2012sensing}. A noise contribution $S(\omega)\sim \lvert\beta_0\rvert^2\delta(\omega-\omega_0)$, for instance, leads for $n=N$ to an amplitude $\tilde{C}(N\pi/\omega_0)\simeq e^{-(N/N_0)^2}$, where $N_0=\sqrt{\pi}\omega_0/(2\lvert \beta_0\rvert)$, giving a visibility $\propto (N/N_0)^2$ that increases with $N$ for $N<N_0$. As $N$ increases, however, the ability to extract information about the qubit coherence may become limited by cavity-induced backaction.

\emph{Backaction}---A protocol that requires continuous monitoring of a qubit in a driven cavity may suffer from backaction induced by qubit dephasing due to cavity-photon shot noise~\cite{gambetta2016qubit}, as well as measurement-induced backaction that necessitates a continuous update of the quantum state \cite{korotkov2016quantum}. In contrast, the protocol presented here involves no direct cavity driving, and the qubit is re-prepared in each measurement cycle. However, coupling to the cavity will still induce unwanted backaction on the qubit through Purcell decay, beyond the minimum backaction required to extract information about the qubit coherence dynamics: For a CPMG sequence, and for $\kappa\tau\gg 1$, we find that inhomogeneously broadened Purcell decay leads to a stretched-exponential decay, $\tilde{C}(n\tau)\propto e^{-\sqrt{\gamma_{\mathrm{P}}n\tau}}$. For $n>1/(\gamma_\mathrm{P}\tau)$, it also gives rise to a simultaneous broadening (in time) and modulation of the echo revivals, 
\begin{equation}
    G_n(t)\simeq e^{-\left(\frac{t}{2T_2^*}\right)^2}\cos\left[\sqrt{2}(\gamma_\mathrm{P}n\tau)^{1/4}\frac{t}{T_2^*}\right],
\end{equation}
leading to an additional suppression of large-$n$ cavity revivals by a factor $\bar{G}_n\simeq 2e^{-2\sqrt{\gamma_\mathrm{P}n\tau}}$ \cite{supplement}. This suppression limits the number of revivals (echoes) that can be measured before coherence decays to zero, and hence, the signal that can be extracted in each cycle from measurements on the transmission line. 

Coherence is transferred from the qubit to the cavity via Eq.~\eqref{eq:cavity-output}, and from the cavity to the output transmission line via $\dot{r}_{k,2}(t)=-i\omega_k r_{k,2}(t)-i\eta_{k,2} a(t)$. After (i) integrating these equations of motion, (ii) tracing out the cavity, qubit, and environment, and (iii) averaging over realizations of the noise $\eta(t)$, we obtain the reduced density matrix $\rho_\mathrm{TL}$ of the output transmission line. Provided there is at most one photon in the transmission line (this limit can always be reached by reducing $\kappa_2/\kappa$),
\begin{equation}
    \rho_{\mathrm{TL}}=(1-S)\rho_{\mathrm{inc}}+S\ket{\psi}\bra{\psi},
\end{equation}
where $\rho_{\mathrm{inc}}$ is the incoherent part of the density matrix [$\mathrm{Tr}(r_{k,2}\rho_{\mathrm{inc}})=0\;\forall k$] and $\ket{\psi}=\frac{1}{\sqrt{2}}(\ket{0}+\ket{1})$. Here, $\ket{1}=2S^{-1}\sum_k\braket{r_{k,2}}r_{k,2}^\dagger\ket{0}$, where $S=2[\sum_k\lvert\braket{r_{k,2}}\rvert^2]^{1/2}$. For $S\to1$, information about the qubit coherence dynamics is fully transferred into the pure state $\ket{\psi}$ of a two-level system, allowing, in principle, up to one bit of information to be extracted per measurement cycle. Typically, however, $S\ll 1$ will be realized, yielding $\ll 1$ bit of information per cycle. For example, we find that a Hahn echo sequence (CPMG with $N=1$) leads to \cite{supplement} 
\begin{equation}
    S\le S_\mathrm{Hahn}=\frac{\sqrt{5\pi}}{2}gT_2^*\left(\frac{\kappa_2}{\kappa}\right)^{1/2},
\end{equation}
limited by $gT_2^*\ll 1$. For a large-$N$ CPMG sequence, by contrast, we find a significantly larger bound, 
\begin{equation}
    S\lesssim S_{\mathrm{CPMG}}= \frac{2\sqrt{\pi}}{3}\left[\left(\frac{\kappa_2}{\kappa}\right)\left(\frac{1}{\kappa\tau}\right)\right]^{1/2},
\end{equation}
still limited by the small parameter $1/\kappa\tau\ll 1$. The CPMG signal is limited because Purcell decay is always active, while coherence is only transferred from the qubit to the transmission line for a small fraction of the time $\sim T_2^*/\tau \ll 1$. Since the times $t=n\tau$ of the revivals are known, we can improve on this limit if the coupling $g=g(t)$ or the detuning $\delta=\delta(t)$ is pulsed to eliminate Purcell decay for $\lvert t-n\tau\rvert\gtrsim T_2^*$. In this case, we find a maximum achievable signal 
\begin{equation}
    S\lesssim S_{\mathrm{max}}=\left(\frac{\kappa_2}{\kappa}\right)^{1/2}
\end{equation}
that approaches $S_\mathrm{max}\simeq 1$ for $\kappa_2\simeq\kappa$ \cite{supplement}. Transient spectroscopy can therefore achieve the same efficiency as a single-shot readout (one bit per cycle).

A central finding of this Letter is an even/odd modulation of echo revivals under dynamical decoupling. This modulation (a unique signature of quantum noise) results from the non-stationary analog of a Lamb shift arising from quantum fluctuations of an environment, an important indicator of non-classical and non-Markovian dynamics \cite{divincenzo2005rigorous,coish2010free}. When the correlation time of the environment is short compared to the typical observation time (Markovian limit), the quantum-noise phase $\Phi_{\mathrm{q}}(t)$ will advance approximately linearly, $\Phi_\mathrm{q}(t)\simeq \Delta\omega_\mathrm{Lamb}t$, reflecting a simple frequency shift. In contrast, for a non-Markovian system, $\Phi_\mathrm{q}(t)$ may have a highly nontrivial time dependence, reflecting a complex quantum dynamics. This phase may be amplified under repeated fast qubit rotations, which have the effect of stroboscopically driving the environment away from stationarity. Ignoring this effect during a quantum computation may then lead to an accumulation of phase errors that could otherwise be fully corrected.

\textit{Note added}---Following submission, we became aware of Ref.~\onlinecite{mutter2022fingerprints}, which considers the influence of low-frequency qubit dephasing noise on the transient cavity transmission. In addition to the free-induction decay considered in Ref.~\onlinecite{mutter2022fingerprints}, we also consider (i) dynamical decoupling sequences applied to the qubit, (ii) quantum noise, (iii) strategies for maximizing the signal, and (iv) cavity-induced backaction (an effect that is higher-order in $g$). As shown here, cavity-induced backaction ultimately sets the limiting timescale (determined by the inhomogeneously broadened Purcell decay time rather than by the cavity decay time $1/\kappa$) for monitoring qubit coherence through the cavity. 

\begin{acknowledgments}
\emph{Acknowledgments}---We thank A.~Blais for useful discussions and  {\L}.~Cywi\'nski for both pointing out an error in our expression for the quantum filter function in an earlier manuscript and for bringing Refs.~\onlinecite{kwiatkowski2020influence, pazsilva2017multiqubit} to our attention. We also acknowledge funding from the Natural Sciences and Engineering Research Council (NSERC) and from the Fonds de Recherche--Nature et Technologies (FRQNT).
\end{acknowledgments}

%

\pagebreak

\providecommand{\noopsort}[1]{}\providecommand{\singleletter}[1]{#1}%

\pagebreak

\setcounter{equation}{0}

\renewcommand{\theequation}{S\arabic{equation}}
\renewcommand{\thefigure}{S\arabic{figure}}
\renewcommand\bibnumfmt[1]{[S#1]}
\renewcommand{\citenumfont}{S}

\setcounter{secnumdepth}{3}
\renewcommand{\thesection}{S\Roman{section}}

\begin{widetext}

\begin{center}
{\large \textbf{Supplemental information for `Non-Markovian transient spectroscopy in cavity QED'}}\\\smallskip
Z. McIntyre and W. A. Coish\\
Department of Physics, McGill University, 3600 rue University, Montreal, QC H3A 2T8, Canada
\end{center} 

\noindent This supplement provides derivations of several results from the main text. In Sec.~\ref{supp-mat-sec:SI-rotating-frame}, we introduce a toggling-frame transformation that accounts for the effects of qubit driving. This toggling frame will be used to perform subsequent calculations. In Sec.~\ref{supp-mat-sec:S2-cavity-field-coherence}, we relate the cavity field to qubit coherence, leading to Eq.~\eqref{eq:cavity-output} of the main text. In Sec.~\ref{supp-mat-sec:S3-qubit-drive}, we analyze the effects of cavity-induced backaction on the qubit and find that inhomogeneously broadened Purcell decay leads to a stretched-exponential decay of the qubit echo envelope as well as a modification of the shape of coherence revivals. We additionally consider qubit coherence under an $N$-pulse Carr-Purcell-Meiboom-Gill (CPMG) dynamical-decoupling sequence, leading to the expression for the cavity field in terms of the CPMG echo envelope presented in Eq.~\eqref{eq:spectrum-peak} of the main text. Section~\ref{supp-mat-sec:S4-quantum-noise} gives explicit expressions for the quantum and classical noise contributions to Eq.~\eqref{eq:spectral-density} of the main text, while Sec.~\ref{supp-mat-sec:s5-input-output} presents the expression for the cavity transmission $A_{\mathrm{T}}(\omega,\eta)$ used to generate Fig.~\ref{fig:inhomogeneous-broadening}(a). Finally, in Sec.~\ref{supp-mat-sec:s6-signal}, we quantify limits on the maximum signal and consider a modified protocol for which the signal is not limited by either inhomogeneous broadening or cavity-induced backaction. In this supplement, boxed equations correspond to displayed equations in the main text.

\section{The toggling frame}
\label{supp-mat-sec:SI-rotating-frame}
\noindent We consider a qubit (``system'') interacting with a quantum environment through the lab-frame Hamiltonian
\begin{equation}
    H_{\mathrm{lab}}(t)=\frac{1}{2}[\Delta+\eta(t)+h]\sigma_z+H_{\mathrm{E}}+H_{\mathrm{drive}}(t),\label{eq-supp-mat:system-environment}
\end{equation}
where $\Delta$ is the bare qubit splitting, $h$ is an operator acting exclusively on the environment (with free Hamiltonian $H_{\mathrm{E}}$), and $H_\mathrm{drive}(t)$ is an arbitrary drive acting only on the qubit [having Pauli-z operator $\sigma_z=\ket{e}\bra{e}-\ket{g}\bra{g}$ for excited (ground) state $\ket{e(g)}$]. Classical dephasing noise is described by $\eta(t)$, which we assume is generated by stationary, zero-mean Gaussian noise fully described by the noise spectrum given in Eq.~\eqref{eq:classical-noise-spectral-density} of the main text,
\begin{equation}
    \boxed{S_\eta(\omega) = \int dt \:e^{-i\omega t}\llangle\eta(t)\eta(0)\rrangle,}
\end{equation}
where $\llangle\cdots\rrangle$ indicates an average over realizations of $\eta(t)$. Inhomogeneous broadening [low-frequency fluctuations in $\eta(t)$] will then lead to Gaussian free-induction decay of the qubit on a time scale $T_2^*$ given by
\begin{equation}
    \frac{2}{\left(T_2^*\right)^2}=\int_{-\infty}^\infty\frac{d\omega}{2\pi}S_\eta(\omega).
\end{equation}

We assume that prior to the start of the experiment (at $t=0$), the environment is allowed to reach a steady-state in contact with the qubit in its ground state. The initial state of the qubit and environment is therefore taken to be of the form $\ket{g}\bra{g}\otimes \bar{\rho}_{\mathrm{E}}$, where $\bar{\rho}_{\mathrm{E}}$ is stationary with respect to the Hamiltonian conditioned on the qubit being in $\ket{g}$:
\begin{equation}
    \commute{H_{\mathrm{E}}-\frac{h}{2}}{\bar{\rho}_{\mathrm{E}}}=0.\label{eq-supp-mat:initial-state}
\end{equation} 
This initial condition, which should be realized generically in experiment, is crucial for our analysis of quantum noise in Sec.~\ref{supp-mat-sec:S4-quantum-noise}, below. The more common ``weak coupling" initial condition, $[H_{\mathrm{E}},\bar{\rho}_{\mathrm{E}}]=0$, yields no quantum noise term for coupling $h\sigma_z/2$~\cite{S-kwiatkowski2020influence, S-norris2016qubit}. 

We treat the deviation from this conditioned Hamiltonian $H_{\mathrm{E}}-h/2$ as a perturbation to the environment dynamics. In an interaction picture defined with respect to $H_{\mathrm{E}}-h/2$, the system-environment Hamiltonian then reads
\begin{align}
    H_{\mathrm{SE}}(t)&=U_{\mathrm{E}}^\dagger(t)H_{\mathrm{lab}}(t)U_{\mathrm{E}}(t)-iU_{\mathrm{E}}^\dagger(t)\dot{U}_{\mathrm{E}}(t),\quad U_{\mathrm{E}}(t)=e^{-i(H_{\mathrm{E}}-h/2)t}\\
    &=\frac{1}{2}[\Delta+\Omega(t)]\sigma_z+\frac{1}{2}h(t)+H_{\mathrm{drive}}(t),\label{eq-supp-mat:system-environment-2}
\end{align}
where we have introduced
\begin{equation}
    \Omega(t)=\eta(t)+h(t).\label{Omega}
\end{equation}

We now consider a master equation for the joint state $\rho(t)$ of the driven qubit and quantum environment, coupled to a cavity mode (with annihilation operator $a$), together with a quasi-continuum of transmission-line modes coupled to the cavity input and output ports, all of which evolve via a time-dependent Hamiltonian $H(t)$. In addition, we assume the qubit has a dephasing rate $\gamma_\phi$ independent of the quantum environment, and that the occupation of the cavity mode decays at an extrinsic rate $\kappa_{\mathrm{ext}}$ independent of coupling to the input and output transmission lines [Eq.~\eqref{eq:master-eq} and Fig.~\ref{fig:cavity-schematic} of the main text]:
\begin{equation}
    \boxed{\dot{\rho}(t)=-i[H(t),\rho(t)]+\frac{\gamma_\phi}{2}\mathcal{D}[\sigma_z]\rho(t)+\kappa_{\mathrm{ext}}\mathcal{D}[a]\rho(t).}\label{eq-supp-mat:master-eq-lab-frame}
\end{equation}
Here, the damping superoperator acts according to $\mathcal{D}[\mathcal{O}]\rho=\mathcal{O}\rho\mathcal{O}^\dagger-\{\mathcal{O}^\dagger\mathcal{O},\rho\}/2$ for an arbitrary operator $\mathcal{O}$, and the Hamiltonian $H(t)$ can be written in terms of the system-environment Hamiltonian [Eq.~\eqref{eq-supp-mat:system-environment-2}] derived above:
\begin{equation}
    \boxed{H(t)=H_{\mathrm{SE}}(t)+\omega_c a^\dagger a +g\sigma_x(a^\dagger+a)+\sum_{i=1,2}\sum_k(\lambda_{k,i}e^{i\omega_k t}r_{k,i}^\dagger a+\mathrm{h.c.}),}\label{eq-supp-mat:lab-frame-hamiltonian}
\end{equation} 
where the qubit couples to the cavity mode with a Rabi coupling of strength $g$, and the cavity mode couples to the transmission-line modes with strengths $\{\lambda_{k,i}\}$. The mode of the input ($i=1$) or output ($i=2$) transmission line having freqency $\omega_k$ is associated with an annihilation operator $r_{k,i}$. 

We now transform to a toggling frame to account for the effect of the qubit drive $H_{\mathrm{drive}}(t)$. Unitary evolution $U(t)$ under the full Hamiltonian $H(t)$ is related to the toggling-frame unitary $\tilde{U}(t)$ through
\begin{equation}
    U(t)=\mathcal{T}e^{-i\int_0^t dt' H(t')}=U_{\mathrm{TF}}(t)\tilde{U}(t),
\end{equation}
where $\mathcal{T}$ is the time-ordering operator, and where
\begin{equation}
    U_{\mathrm{TF}}(t)=U_{\mathrm{drive}}(t)R(t).
\end{equation}
Here, $U_\mathrm{drive}(t)$ eliminates evolution under $H_\mathrm{drive}(t)$,
\begin{equation}
    U_{\mathrm{drive}}(t)= \mathcal{T}e^{-i\int_0^tdt'H_{\mathrm{drive}}(t')},
\end{equation}
and $R(t)$ defines the rotating frame subject to $U_\mathrm{drive}(t)$:
\begin{equation}
    R(t)=\mathcal{T}e^{-i\Delta\int_0^t dt'[a^\dagger a+U_{\mathrm{drive}}^\dagger (t')\sigma_z U_{\mathrm{drive}}(t')/2]}.
\end{equation}
This transformation allows for a simpler analysis of observables $\tilde{\mathcal{O}}(t)$ evolving under the action of the toggling-frame Hamiltonian $\tilde{H}(t)$:
\begin{eqnarray}
    \tilde{H}(t) & = & U_{\mathrm{TF}}^\dagger(t) H(t) U_{\mathrm{TF}}(t) -iU_{\mathrm{TF}}^\dagger(t)\dot{U}_{\mathrm{TF}}(t),\\
    \tilde{\mathcal{O}}(t) & = & \tilde{U}^\dagger(t)\mathcal{O}\tilde{U}(t),\quad \tilde{U}(t)=\mathcal{T}e^{-i\int_0^t dt'\tilde{H}(t')}.
\end{eqnarray}
The expectation value $\braket{\mathcal{O}}_t$ can then be related to $\braket{\tilde{\mathcal{O}}}_t$ through
\begin{equation}
    \braket{\mathcal{O}}_t=\llangle\mathrm{Tr}\{U^\dagger(t)\mathcal{O}U(t)\rho(0)\}\rrangle=\llangle\mathrm{Tr}\{\hat{\mathcal{O}}(t)\tilde{\rho}(t)\}\rrangle;\quad \tilde{\rho}(t)=\tilde{U}(t)\rho(0)\tilde{U}^\dagger(t),\label{eq-supp-mat:expectation-val}
\end{equation}
where we have included both the quantum average $\mathrm{Tr}\left\{\cdots \rho(0)\right\}$ and the classical average over noise realizations $\llangle\cdots \rrangle$ in the definition of the expectation value $\braket{\cdots}_t$. Further, we denote by a ``hat'' the analog of an interaction-picture operator:
\begin{equation}
    \hat{\mathcal{O}}(t)=U_{\mathrm{TF}}^\dagger(t) \mathcal{O} U_{\mathrm{TF}}(t).\label{eq-supp-mat:toggling-frame-op}
\end{equation}
These definitions give, for example, $\braket{a}_t=\llangle\mathrm{Tr}\{\hat{a}(t)\tilde{\rho}(t)\}\rrangle=e^{-i\Delta t}\llangle\mathrm{Tr}\{a\tilde{\rho}(t)\}\rrangle=e^{-i\Delta t}\braket{\tilde{a}}_t$. The toggling-frame density operator evolves under
\begin{equation}
    \dot{\tilde{\rho}}(t)=-i[\tilde{H}(t),\tilde{\rho}(t)]+\frac{\gamma_\phi}{2}\mathcal{D}[\hat{\sigma}_z(t)]\tilde{\rho}(t)+\kappa_{\mathrm{ext}}\mathcal{D}[a]\tilde{\rho}(t),\label{eq-supp-mat:master-eq-rot-frame}
\end{equation}
where, in writing the transformed damping superoperators, we have used $\hat{\sigma}_z^2(t)=\sigma_z^2=1$ and $\hat{a}(t)=e^{i\Delta t}a$. In terms of the cavity-qubit detuning $\delta=\omega_c-\Delta$, the toggling-frame Hamiltonian is now given by
\begin{equation}
    \tilde{H}(t)=\tilde{H}_{\mathrm{SE}}(t)+\delta a^\dagger a+g\hat{\sigma}_x(t)[e^{i\Delta t}a^\dagger+\mathrm{h.c.}]+\sum_{i=1,2}\sum_k(\lambda_{k,i}e^{i(\omega_k-\Delta) t}r_{k,i}^\dagger a+\mathrm{h.c.}),\label{eq-supp-mat:hamiltonian}
\end{equation}
where
\begin{equation}
    \tilde{H}_{\mathrm{SE}}(t)=\frac{1}{2}\Omega(t)\hat{\sigma}_z(t)+\frac{1}{2}h(t).\label{eq-supp-mat:toggling-ham-0}
\end{equation}

\section{Relating the output field to qubit coherence}
\label{supp-mat-sec:S2-cavity-field-coherence}

\noindent We can recover the well-known input-output relation~\cite{S-gardiner1985input} by integrating the Heisenberg equation of motion, $\dot{r}_{k,i}(t)=i\commute{H(t)}{r_{k,i}(t)}$, resulting in
\begin{equation}
    \braket{r_{k,i}}_t=e^{-i\omega_kt} \braket{r_{k,i}}_0-i\lambda_{k,i}\int_0^t dt'\:e^{-i\omega_k(t-t')}e^{-i\Delta t'}\braket{\tilde{a}}_{t'},\quad i=1,2.\label{eq-supp-mat:eom-rk}
\end{equation}
Summing Eq.~\eqref{eq-supp-mat:eom-rk} over a quasi-continuous set of modes $k$ and performing a Markov approximation for wide-bandwidth transmission lines gives the input-output relation
\begin{equation}
    r_{\mathrm{out},i}(t) = r_{\mathrm{in},i}(t)-i\sqrt{\kappa_i}e^{-i\Delta t}\braket{\tilde{a}}_t,\label{eq-supp-mat:input-output}
\end{equation}
where
\begin{equation}
    r_{\mathrm{out},i}(t)=\sqrt{\frac{c}{L}}\sum_k \braket{r_{k,i}}_t,\quad r_{\mathrm{in},i}(t)=\sqrt{\frac{c}{L}}\sum_ke^{-i\omega_kt}\braket{r_{k,i}}_0, \quad \kappa_i=\frac{L}{c}\lvert\lambda_i(\omega_c)\rvert^2
\end{equation}
for $\lambda_i(\omega=\omega_k)=\lambda_{k,i}$. Here, we have assumed one-dimensional transmission lines of length $L$ supporting linearly-dispersing modes ($\omega_k=c|k|$) with speed of light $c$. In order to relate the transmission-line dynamics more transparently to qubit coherence dynamics, we consider the quantum Langevin equation for the cavity field $\tilde{a}(t)$. Within the same Markov approximation used to obtain Eq.~\eqref{eq-supp-mat:input-output}, this equation reads
\begin{equation}
    \dot{\tilde{a}}(t)=-\left(i\delta+\frac{\kappa}{2}\right)\tilde{a}(t)-ige^{i\Delta t}\sigma_x(t),\label{eq-supp-mat:eom-cavity}
\end{equation}
where $\kappa=\kappa_1+\kappa_2+\kappa_{\mathrm{ext}}.$ Equation \eqref{eq-supp-mat:eom-cavity} provides a useful relation between $\sigma_x(t)$ (the lab-frame qubit coherence) and the cavity field $\tilde{a}(t)=e^{i\Delta t}a(t)$. This relationship is valid for an arbitrary qubit drive and for arbitrarily large qubit-cavity coupling $g$. 

The goal is now to understand dynamics of the cavity field evolving under Eq.~\eqref{eq-supp-mat:eom-cavity} due to some specific driven qubit dynamics $\sigma_x(t)$. As explained in Sec.~\ref{supp-mat-sec:SI-rotating-frame}, above, we assume that the joint state of the qubit, cavity, environment, and transmission line is given, for $t\le 0$, by $\rho(t\le 0)=\ket{g,0,0}\bra{g,0,0}\otimes\bar{\rho}_{\mathrm{E}}$, where $\ket{\sigma,n_c,\nu}$ denotes the state of the qubit ($\sigma=g,e$), the cavity mode containing $n_c$ photons, and the transmission line ($\nu=0$ is the vacuum for all $i,k$). The initial state of the environment, $\rho_{\mathrm{E}}(t\le 0)=\bar{\rho}_{\mathrm{E}}$, is assumed to be stationary for $t<0$. This is true provided (i) the qubit drive is not turned on until $t=0$, $H_{\mathrm{drive}}(t<0)=0$, (ii) the qubit-environment interaction $h\sigma_z/2$ is secular ($\commute{h\sigma_z}{\sigma_z}=0$, as assumed above), and (iii) the environment has reached a steady-state in contact with the qubit, $[H_{\mathrm{E}}-h/2,\bar{\rho}_{\mathrm{E}}]=0$ [Eq.~\eqref{eq-supp-mat:initial-state}]. The state $\ket{g,0,0}$ is stationary provided either the qubit-cavity coupling vanishes [$g(t)=0$ for $t<0$], or the qubit and cavity are far detuned until $t=0$, so that $\ket{g,0,0}$ is an eigenstate of $H(t)$ with small corrections. Integrating and averaging Eq.~\eqref{eq-supp-mat:eom-cavity} under these assumptions recovers Eq.~\eqref{eq:cavity-output} of the main text,
\begin{equation}
    \boxed{\braket{\tilde{a}}_t=-ig\int_{-\infty}^\infty dt'\:\chi_c(t-t')e^{i\Delta t'}\braket{\sigma_x}_{t'},\quad\chi_c(t)=e^{-i\delta t-\tfrac{\kappa}{2}t}\Theta(t),\quad\braket{\sigma_x}_t\propto\Theta(t),}\label{eq-supp-mat:cavity-output}
\end{equation}
where $\Theta(t)$ is a Heaviside function. A finite qubit coherence, $\braket{\sigma_x}_t\neq 0$, can be introduced at $t=0$ with, e.g., a rapid $\pi/2$-pulse, after which the cavity field will evolve according to Eq.~\eqref{eq-supp-mat:cavity-output}.

In terms of the Fourier transform, $\braket{\mathcal{O}}_\omega=\int dt\:e^{i\omega t}\braket{\mathcal{O}}_t$, Eq.~\eqref{eq-supp-mat:cavity-output} reads
\begin{equation}
    \braket{\tilde{a}}_\omega=-ig\chi_c(\omega)\braket{\sigma_x}_{\omega+\Delta},\quad\chi_c(\omega)=\int_{-\infty}^\infty dt\: e^{i\omega t}\chi_c(t)=\frac{1}{i(\delta-\omega)+\kappa/2}.\label{eq-supp-mat:cavity-output-freq}
\end{equation}
In the limit of low $Q=\omega_c/\kappa<1$, we can take $\chi_c(\omega)\sim 2/\kappa$ to be flat on the scale of variation of $\braket{\sigma_x}_{\omega+\Delta}$ for $\Delta\sim\omega_c$. The cavity field consequently mirrors the dynamics of the qubit time-locally: $\braket{a}_t\propto\braket{\sigma_x}_t$ \cite{S-ranjan2020pulsed}. Notably, Eqs.~\eqref{eq-supp-mat:cavity-output} and \eqref{eq-supp-mat:cavity-output-freq} accurately reflect dynamics even in the regime of high $Q$. This is typically the regime of interest for the devices (see, e.g.,~\cite{S-stockklauser2017strong, S-mi2017strong, S-mi2018coherent, S-samkharadze2018strong,S-landig2018coherent, S-cubaynes2019highly, S-viennot2015coherent}) designed to reach the strong-coupling regime of cavity QED. The high-$Q$ regime also admits a cavity-filter approximation, in which we replace $\braket{\sigma_x}_t\simeq\braket{\sigma_-}_t$ in the convolution:
\begin{equation}
    \braket{\tilde{a}}_t\simeq-ig\int_{-\infty}^\infty dt'\:\chi_c(t-t')e^{i\Delta t'}\braket{\sigma_-}_{t'}\quad \left[\mathrm{high-}Q:\;\mathrm{max}\left(|\delta|,\kappa\right)\ll|\Delta|\right].\label{eq-supp-mat:cavity-output-high-Q}
\end{equation}
A direct consequence of this cavity-filter (high-$Q$) approximation is that the cavity field $\braket{\tilde{a}}_t$ [and hence, the output field $r_\mathrm{out,2}(t)$ via Eq.~\eqref{eq-supp-mat:input-output}] will show a unique signature of quantum noise when we consider a dynamical decoupling sequence in Sec.~\ref{supp-mat-sec:S3-qubit-drive}, below.

Equation \eqref{eq-supp-mat:cavity-output-high-Q}, together with the input-output relation, Eq.~\eqref{eq-supp-mat:input-output}, provides a direct link between the dynamics of the output field $r_{\mathrm{out},2}(t)$ and qubit coherence dynamics $\braket{\sigma_-}_t$. No stationarity assumption, weak-coupling approximation, or Markov approximation has been made on $\braket{\sigma_-}_t$ up to this point. Provided an accurate model of non-Markovian dynamics can be found for $\braket{\sigma_-}_t$ (under, say, a dynamical decoupling sequence), this model can be directly tested from a measurement of $r_\mathrm{out,2}(t)$. Alternatively, non-Markovian dynamics in $\braket{\sigma_-}_t$ can be inferred from the transient dynamics of $r_{\mathrm{out},2}(t)$. The qubit dynamics translated to the output field will, however, depend on the effects of the cavity filter and cavity-induced backaction, as we now show.

\section{Cavity-induced backaction}
\label{supp-mat-sec:S3-qubit-drive}

\noindent We assume that qubit coherence is created at $t=0$ with a rapid $({-}\pi/2)_y$-rotation: $\rho(0^+)=\ketbra{+}{+}\otimes\bar{\rho}_{\mathrm{E}}\otimes\ketbra{0,0}{0,0}$, where $\sigma_x\ket{+}=\ket{+}$, and where $\ket{0,0}$ indicates the simultaneous vacuum state of the cavity and transmission line. (We use the notation $\theta_\alpha$ to indicate an infinitesimal-duration rotation of the qubit by angle $\theta$ about the $\alpha$-axis.) The initial $\pi/2$-pulse is followed by a sequence of dynamical-decoupling $\pi_x$-pulses due to $H_{\mathrm{drive}}(t)$.  For such a pulse sequence, $\hat{\sigma}_z(t)=U_{\mathrm{drive}}^\dagger(t)\sigma_zU_{\mathrm{drive}}(t)=s(t)\sigma_z$, where $s(t)=(-1)^{n(t)}$ is a sign function depending on $n(t)$, the number of $\pi$-pulses having taken place up to time $t$. The lab-frame expectation value $\braket{\sigma_-}_t$ in Eq.~\eqref{eq-supp-mat:cavity-output-high-Q} is then related to toggling-frame observables via:
\begin{equation}
    \braket{\sigma_-}_t = \begin{cases}
    e^{-i\phi(t)}\braket{\tilde{\sigma}_-}_t,\quad &n(t)\text{ even}\\
    e^{i\phi(t)}\braket{\tilde{\sigma}_+}_t,\quad &n(t)\text{ odd}
    \end{cases};\quad \phi(t)=\int_0^t dt' s(t')\Delta.\label{eq-supp-mat:lab-toggle}
\end{equation}
The phase $\phi(t)$ advances at a rate $\dot{\phi}(t)=+\Delta$ for $n(t)$ even and $\dot{\phi}(t)=-\Delta$ for $n(t)$ odd, so $\braket{\sigma_-}_t\sim e^{-i\Delta t}$ for all times $t$ up to corrections $\sim\braket{\tilde{\sigma}_\pm}_t$. To solve for $\braket{\tilde{\sigma}_\pm}_t$, we evaluate the Heisenberg equations of motion under $\tilde{H}(t)$ [from Eq.~\eqref{eq-supp-mat:hamiltonian}], which we rewrite as 
\begin{equation}
    \tilde{H}(t)=\tilde{H}_{\mathrm{SE}}(t)+\delta a^\dagger a+\sum_{i=1,2}\sum_k(\lambda_{k,i}e^{i(\omega_k-\Delta) t}r_{k,i}^\dagger a+\mathrm{h.c.})+\begin{cases}
    g(e^{i(\phi(t)-\Delta t)}\sigma_+a+\mathrm{h.c.})+\text{c.r.},\quad &n(t)\text{ even}\\
    g(e^{i(\phi(t)+\Delta t)}\sigma_+a^\dagger+\mathrm{h.c.})+\text{c.r.},\quad &n(t)\text{ odd}
    \end{cases},\label{eq-supp-mat:htilde}
\end{equation}
where
\begin{equation}
    \tilde{H}_{\mathrm{SE}}(t)=\frac{1}{2}\Omega(t)s(t)\sigma_z+\frac{1}{2}h(t);\quad \Omega(t)=\eta(t)+h(t).\label{eq-supp-mat:htilde-SE}
\end{equation}
In Eq.~\eqref{eq-supp-mat:htilde}, ``c.r.''~indicates counter-rotating terms $\sim e^{\pm i2\Delta t}$ that lead to small corrections for $|g|\ll|\delta\pm\Delta|$. The usual excitation-preserving co-rotating terms ($\sim \sigma_+a$ and $\sim\sigma_-a^\dagger$) for $n(t)$ even are replaced by excitation non-conserving terms ($\sim \sigma_-a$ and $\sigma_+a^\dagger$) for $n(t)$ odd. These will generally lead to cavity heating, making the analysis of qubit-cavity dynamics under a dynamical decoupling sequence more challenging than for the undriven case \cite{S-beaudoin2017hamiltonian}. These effects can nevertheless be controlled in the appropriate limits. Neglecting counter-rotating terms in $\tilde{H}(t)$, the equation of motion for $\tilde{\sigma}_-(t)$ is then given [within the rotating-wave approximation (RWA)] by
\begin{equation}
    \dot{\tilde{\sigma}}_-(t)\simeq i[\tilde{H}_{\mathrm{SE}}(t),\tilde{\sigma}_-(t)]-\gamma_\phi\tilde{\sigma}_-(t)+\begin{cases}
    ig e^{i\left[\phi(t)-\Delta t\right]}\tilde{\sigma}_z(t)\tilde{a}(t),\quad &n(t)\text{ even}\\
    ig e^{i\left[\phi(t)+\Delta t\right]}\tilde{\sigma}_z(t)\tilde{a}^\dagger(t),\quad &n(t)\text{ odd}
    \end{cases},\quad \left(\mathrm{RWA:}\, g\ll |\delta\pm\Delta|\right).\label{eq-supp-mat:sigma-minus-eom}
\end{equation}
The bilinear terms $\sim \tilde{\sigma}_z(t)\tilde{a}(t)$ and $\sim \tilde{\sigma}_z(t)\tilde{a}^\dagger(t)$ make a general integration of these equations difficult. However, we note that for free-induction decay [$n(t)=0$ for all $ t$], the dynamics are restricted (under the rotating-wave approximation and for an undriven cavity with $\kappa_1=0$) to the subspace spanned by $\left\{\ket{g,0,0},\ket{e,0,0},\ket{g,1,0},\ket{g,0,k}\right\}$, where $\ket{g,0,k}=r_{k,2}^\dagger\ket{g,0,0}$. In this case, the state of the qubit and cavity is restricted to the bottom three rungs of the Jaynes-Cummings ladder.  Within this subspace, we have $\left<\tilde{\sigma}_z(t)\tilde{a}(t)\right>=-\braket{\tilde{a}}_t$ for all time, allowing for a direct solution to the coupled equations for $\braket{\tilde{\sigma}_-}_t$ and $\braket{\tilde{a}}_t$. As described above, this exact replacement is no longer possible under a dynamical decoupling sequence. However, we can still justify a similar approximate replacement provided (i) that $g\ll\kappa$, so that the cavity contains at most one photon at any time, and (ii) that the minimum time $\tau$ between $\pi$-pulses is long compared to the timescale $\kappa^{-1}$ of cavity transients. Under these conditions, we perform the restricted-subspace approximation:
\begin{equation}
    \braket{\tilde{\sigma}_z(t)\tilde{a}(t)}\simeq -\braket{\tilde{a}(t)},\quad n(t)\text{ even};\quad
    \braket{\tilde{\sigma}_z(t)\tilde{a}^\dagger(t)}\simeq \braket{\tilde{a}^\dagger(t)},\quad n(t)\text{ odd};\quad (g<\kappa,\,\kappa\tau\gg 1).\label{eq-supp-mat:subspace}
\end{equation}
The first approximate equality follows from the same logic given above for free-induction decay, $n(t)=0$. The second approximation follows provided evolution is approximately restricted to the subspace $\left\{\ket{g,0,0},\ket{e,0,0},\ket{e,1,0},\ket{e,0,k}\right\}$ for most of the time when $n(t)$ is odd. These approximations will be violated due to heating effects on a time scale $\sim \kappa^{-1}$ in the vicinity of $\pi$-pulses, but if the time between subsequent $\pi$-pulses is sufficiently long, the cumulative effect of these transients will amount to a small correction to the qubit coherence dynamics. 
\begin{figure}
    \centering
    \includegraphics[width=0.8\linewidth]{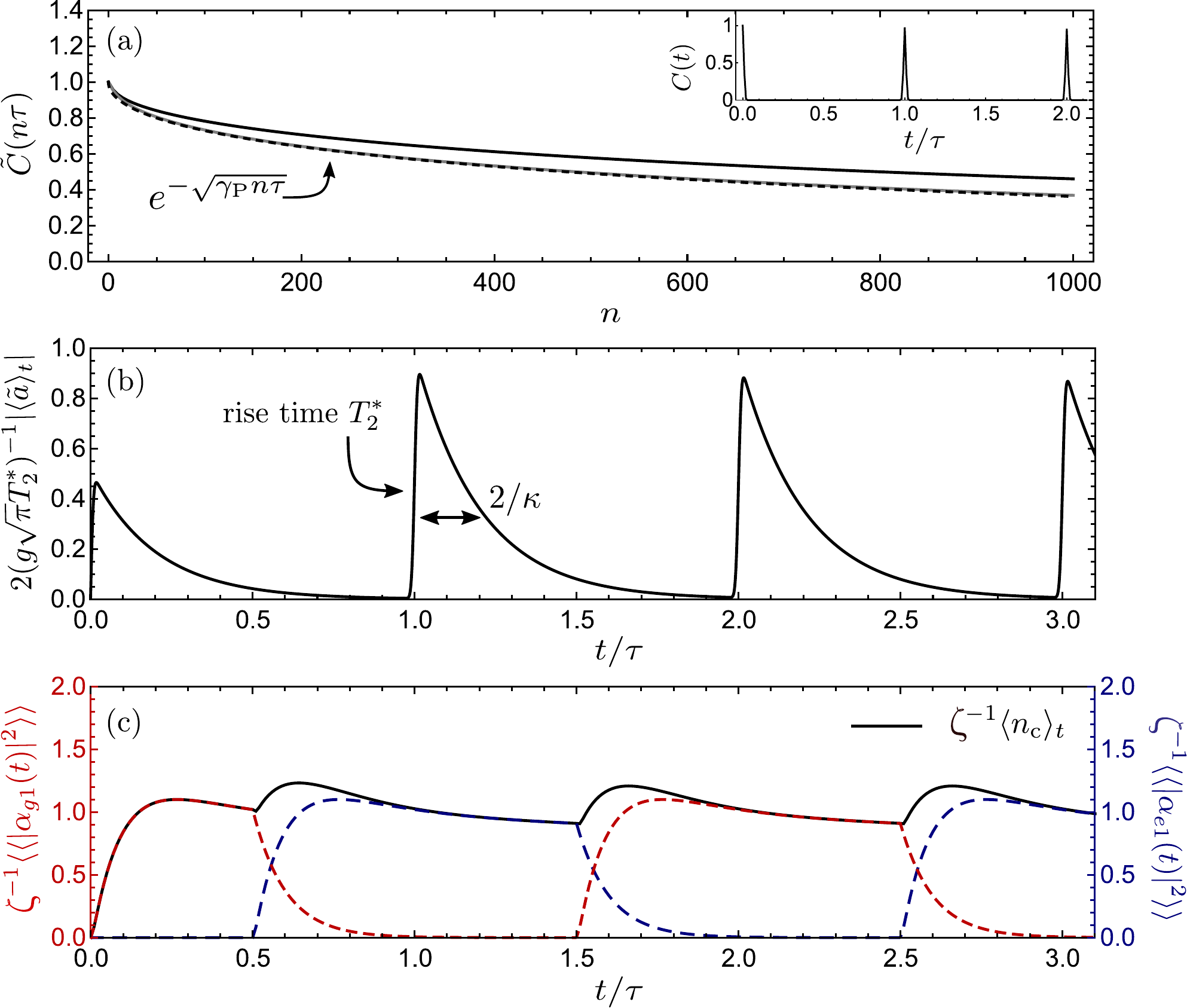}
    \caption{(a) Solid black line: The echo envelope $\tilde{C}(n\tau)=\left\llangle\braket{\tilde{\psi}(n\tau)\lvert\sigma_-\rvert\tilde{\psi}(n\tau)}\right\rrangle/\braket{\sigma_-}_0$, where  $\ket{\tilde{\psi}}$ evolves under $\tilde{H}(t)$ in the restricted subspace described by Eq.~\eqref{eq-supp-mat:basis}. Gray line: $\tilde{C}(n\tau)\simeq\llangle e^{-\Gamma_{\mathrm{P}}(\eta)n\tau/2}\rrangle$. Black dashed line: The approximate form $\tilde{C}(n\tau)\simeq e^{-\sqrt{\gamma_{\mathrm{P}}n\tau}}$. Inset: The qubit coherence $C(t)$, consisting of Gaussian revivals with width $\sim T_2^*$ centered at times $t=n\tau$, with $n$ an integer. (b) The cavity field $\braket{\tilde{a}}_t=\sum_{\sigma=e,g}\left\llangle\alpha_{\sigma 0}^*(t)\alpha_{\sigma 1}(t)\right\rrangle$. (c) Red (blue) dashed line: $\left\llangle|\alpha_{g1}(t)|^2\right\rrangle$ [$\left\llangle|\alpha_{e1}(t)|^2\right\rrangle$], normalized by $\zeta=\sqrt{\pi}g^2T_2^*/\kappa$. Black line: The cavity occupation $\braket{n_c}_t=\sum_{\sigma=e,g}\left\llangle\lvert\alpha_{\sigma 1}(t)\rvert^2\right\rrangle$. We have verified that the inequality $\lvert\braket{\tilde{a}}_t\rvert^2\leq\braket{n_c}_t(1-\braket{n_c}_t)$ is satisfied for all times $t$, as required by positivity of the cavity density matrix in the subspace of $n_c=0,1$. (Though unresolvable on this scale, $\braket{n_c}_t$ also rises initially on a timescale $T_2^*$.) For all figures, we take $g=0.1\kappa$, $\kappa T_2^*=0.1$ (giving $\sqrt{\pi}gT_2^* \sim 10^{-2}$, $\zeta \sim 10^{-3}$), and $\kappa\tau=10$.}
    \label{fig:backaction}
\end{figure}

To illustrate the validity of the restricted-subspace approximation, Eq.~\eqref{eq-supp-mat:subspace}, we consider the simplified case of $\gamma_\phi=h(t)=0$ and additionally assume that the low-frequency noise is static [$\eta(t)=\eta$]. We then directly integrate the Schr\"odinger equation [$\partial_{t}\ket{\tilde{\psi}(t)}=-i\tilde{H}(t)\ket{\tilde{\psi}(t)}$] \emph{without} assuming Eq.~\eqref{eq-supp-mat:subspace}, but restricting to a subspace that allows for at most one photon in the cavity or transmission line (this can always be justified for $g<\kappa$ at a sufficiently short time):
\begin{equation}
    \ket{\tilde{\psi}(t)}=\alpha_{g0}(t)\ket{g,0,0}+\alpha_{e0}(t)\ket{e,0,0}+\alpha_{g1}(t)\ket{g,1,0}+\alpha_{e1}(t)\ket{e,1,0}+\sum_k\left[\alpha_{gk}(t)\ket{g,0,k}+\alpha_{ek}(t)\ket{e,0,k}\right].\label{eq-supp-mat:basis}
\end{equation}
For a fixed value of the noise parameter $\eta$, the Schr\"odinger equation can be integrated analytically piecewise for each time interval between $\pi$-pulses. After averaging over a Gaussian distribution in $\eta$ values with $\llangle\eta^2\rrangle=2/(T_2^*)^2$, the resulting contributions $\llangle|\alpha_{g1}(t)|^2\rrangle$ and $\llangle|\alpha_{e1}(t)|^2\rrangle$ to the number of cavity photons, $\braket{n_c}_t=\sum_{\sigma={e,g}}\llangle|\alpha_{\sigma 1}(t)|^2\rrangle$, are shown in Fig.~\ref{fig:backaction}(c) for the case of a Carr-Purcell-Meiboom-Gill (CPMG) sequence:
\begin{equation}
    \left(\frac{\tau}{2}-\pi_x-\frac{\tau}{2}\right)^N\quad \mathrm{(CPMG)}.\label{eq-supp-mat:CPMG}
\end{equation} 
Here, $\tau/2$ denotes a delay of duration $\tau/2$. Up to small transient corrections on a time scale $\sim 1/\kappa$ around the times of the $\pi$-pulses, we have $\alpha_{e1}(t)\simeq 0$ for $n(t)$ even and $\alpha_{g1}(t)\simeq 0$ for $n(t)$ odd, justifying the approximation in Eq.~\eqref{eq-supp-mat:subspace}. In addition, the exact echo envelope for this case, $\tilde{C}(n\tau)=\left\llangle\bra{\tilde{\psi}(n\tau)}\sigma_-\ket{\tilde{\psi}(n\tau)}\right\rrangle/\braket{\sigma_-}_0$, is shown in Fig.~\ref{fig:backaction}(a), and the cavity field  $\braket{\tilde{a}}_t=\left\llangle\bra{\tilde{\psi}(t)}a\ket{\tilde{\psi}(t)}\right\rrangle$ is shown in Fig.~\ref{fig:backaction}(b). Even in this case, where there is no external source of pure-dephasing dynamics [$\gamma_\phi=h(t)=0$ and $\eta(t)=\eta$], qubit coherence is still lost due to inhomogeneously broadened Purcell decay, an effect that we now consider in detail.

To separate fast and slow dynamics, we write $\Omega(t)=\eta+\delta\Omega(t)$, where we assume that $\eta$ is a large static contribution (inhomogeneous broadening) and that $\delta\Omega(t)=\delta \eta(t)+h(t)$ generates pure-dephasing dynamics that are slow on the scale $\kappa^{-1}$. We now eliminate the evolution under the fast term $\sim s(t)\eta$ by defining 
\begin{equation}
    \dbtilde{\sigma}_-(t)=e^{i\int_0^tdt's(t')\eta}\tilde{\sigma}_-(t).\label{eq-supp-mat:slow-ev}
\end{equation}
In terms of this quantity, and within the restricted subspace approximation [Eq.~\eqref{eq-supp-mat:subspace}], Eq.~\eqref{eq-supp-mat:sigma-minus-eom} can be written as 
\begin{equation}
    \dot{\dbtilde{\sigma}}_-(t)\simeq i[\tilde{V}(t),\dbtilde{\sigma}_-(t)]-\gamma_\phi\dbtilde{\sigma}_-(t) +ig e^{i\int_0^tdt's(t')\eta}\begin{cases}
    {-}e^{i[\phi(t)-\Delta t]}\tilde{a}(t),\quad &n(t)\text{ even}\\
    e^{i[\phi(t)+\Delta t]}\tilde{a}^\dagger(t),\quad &n(t)\text{ odd}
    \end{cases},\label{eq-supp-mat:sigma-minus-eom-2}
\end{equation}
where
\begin{equation}
    \tilde{V}(t)=\tilde{H}_{\mathrm{SE}}(t)-\frac{1}{2}\eta s(t)\sigma_z=\frac{1}{2}\delta\Omega(t)s(t)\sigma_z+\frac{1}{2}h(t).
\end{equation}
In order to rewrite Eq.~\eqref{eq-supp-mat:sigma-minus-eom-2} as a closed equation, we insert the result for $\tilde{a}(t)$ in terms of $\sigma_-(t)$ within the cavity-filter (high-$Q$) approximation [leading to Eq.~\eqref{eq-supp-mat:cavity-output-high-Q} after averaging]. Neglecting contributions $\sim a(0),\,r_{k,i}(0)$ that vanish under the average ($\braket{\tilde{a}}_0=\braket{r_{k,i}}_0=0$), this gives:
\begin{equation}
\dot{\dbtilde{\sigma}}_-(t)  \simeq i[\tilde{V}(t),\dbtilde{\sigma}_-(t)]-\gamma_\phi\dbtilde{\sigma}_-(t)  -i\int_0^t dt'\Sigma(t,t')\dbtilde{\sigma}_-(t'),\label{eq-supp-mat:sigma-minus-eom-3}
\end{equation}
with a time-nonlocal memory kernel (self-energy) given by 
\begin{equation}
    \Sigma(t,t') = -ig^2 e^{i\int_{t'}^tdt''s(t'')\eta}\begin{cases}
    \chi_c(t-t')e^{i[\phi(t)-\phi(t')-\Delta(t-t')]},\quad &n(t)\text{ even}\\
    \chi_c^*(t-t')e^{i[\phi(t)-\phi(t')+\Delta(t-t')]},\quad &n(t)\text{ odd}
    \end{cases}.\label{eq-supp-mat:self-energy}
\end{equation}
The cavity susceptibility $\chi_c(t-t')$ suppresses contributions for which the times $t,t'$ are well separated; major contributions to the integral thus occur for $t-t' \lesssim \kappa^{-1}\ll \tau$. Except for small intervals of width $\sim 1/\kappa$ around the time of the $\pi$-pulses, we thus have $n(t)=n(t')$ wherever $\chi_c(t-t')$ has significant weight, giving $\phi(t)-\phi(t')\simeq s(t)\Delta(t-t')$. This justifies the following replacements, with small corrections for $\kappa\tau\gg 1$:
\begin{equation}
e^{i[\phi(t)-\phi(t')-\Delta(t-t')]}  \simeq  1,\quad [n(t)\,\mathrm{even}];\quad e^{i[\phi(t)-\phi(t')+\Delta(t-t')]}  \simeq  1,\quad [n(t)\,\mathrm{odd}],\quad\quad (\kappa\tau\gg 1).
\end{equation} 
With these replacements, the self-energy becomes
\begin{equation}
    \Sigma(t,t')\simeq \Sigma_0(t,t-t'),\quad \Sigma_0(t_1,t_2)= -ig^2e^{-\frac{\kappa}{2}t_2}e^{is(t_1)(\eta-\delta)t_2},\quad (\kappa\tau\gg 1).
\end{equation}
If $\dbtilde{\sigma}_-(t)$ evolves slowly on the timescale $\kappa^{-1}$, then we can write the equation of motion for $\dbtilde{\sigma}_-$ in terms of the dispersive shift $\Delta\omega(\eta)$ and Purcell decay rate $\Gamma_P(\eta)$ as
\begin{equation}
    \dot{\dbtilde{\sigma}}_-(t)\simeq i[\tilde{V}(t),\dbtilde{\sigma}_-(t)] -\left\{is(t)\Delta\omega(\eta)+\gamma_\phi+\frac{1}{2}\Gamma_{\mathrm{P}}(\eta)\right\}\dbtilde{\sigma}_-(t),\label{eq-supp-mat:sigma-minus-eom-4}
\end{equation}
where 
\begin{equation}
    i\int_0^\infty dt'\Sigma_0(t,t')=\frac{g^2\left[\kappa/2+is(t)(\eta-\delta)\right]}{(\eta-\delta)^2+(\kappa/2)^2}=\frac{1}{2}\Gamma_{\mathrm{P}}(\eta)+is(t)\Delta\omega(\eta).
\end{equation}

Integrating Eq.~\eqref{eq-supp-mat:sigma-minus-eom-4} and transforming back to $\tilde{\sigma}_-(t)$ [via Eq.~\eqref{eq-supp-mat:slow-ev}] then gives
\begin{align}
    \braket{\tilde{\sigma}_-}_ t\simeq e^{-\gamma_\phi t}\braket{e^{-\frac{\Gamma_{\mathrm{P}}(\eta)}{2}t}e^{-i\int_0^tdt's(t')[\Delta\omega(\eta)+\eta]} \tilde{U}_-^\dagger(t)\tilde{U}_+(t)}\braket{\sigma_-}_0,\label{eq-supp-mat:sigma-minus}
\end{align}
where 
\begin{equation}
    \tilde{U}_\pm(t)=\mathcal{T}e^{-i\int_0^t dt' \tilde{V}_\pm(t')},\quad \tilde{V}_\pm(t) = \braket{e,g\lvert \tilde{V}(t)\rvert e,g}=\frac{1}{2}[h(t)\pm \delta\Omega(t)s(t)].\label{eq-supp-mat:conditional-unitaries}
\end{equation}
As explained in Sec.~\ref{supp-mat-sec:SI-rotating-frame}, $\left<\cdots\right>$ includes both a quantum average and an average over realizations of the noise $\eta(t)$. If we assume that the inhomogeneous broadening $\eta$ is approximately statistically independent of the dynamical contribution $\delta\eta(t)=\eta(t)-\eta$ over the short timescale $\sim T_2^*$ of the coherence revivals, $\llangle\delta\eta(t)\eta\rrangle\simeq 0$, then we can write
\begin{align}
    \braket{\tilde{\sigma}_-}_ t &\simeq \left\llangle e^{-\frac{\Gamma_{\mathrm{P}}(\eta)}{2}t}e^{-i\int_0^tdt's(t')[\Delta\omega(\eta)+\eta]}\right\rrangle \tilde{C}_0(t)\braket{\sigma_-}_0,\label{eq-supp-mat:sigma-minus-2}\\
    \tilde{C}_0(t)&=e^{-\gamma_\phi t} \braket{\tilde{U}_-^\dagger (t)\tilde{U}_+(t)},\label{eq-supp-mat:C0}
\end{align}
where $\tilde{C}_0(t)$ describes the contribution to the slowly varying envelope of qubit coherence due to the environment and low-frequency noise, in the absence of cavity coupling. 

Provided the restricted-subspace approximation [Eq.~\eqref{eq-supp-mat:subspace}] holds, the result given in Eqs.~\eqref{eq-supp-mat:sigma-minus-2} and \eqref{eq-supp-mat:C0} can be used to describe qubit coherence dynamics under an arbitrary dynamical decoupling sequence. For concreteness, we now specialize to an $N$-pulse CPMG sequence with pulse spacing $\tau$ [Eq.~\eqref{eq-supp-mat:CPMG}]. In this case, $\int_0^{n\tau}dts(t)=0$, leading to a perfect cancellation of the inhomogeneous broadening $\eta$ and dispersive shift $\Delta\omega(\eta)$ at revival/echo times $t=n\tau$. These echoes are suppressed by an overall decay envelope set by $\llangle e^{-\Gamma_{\mathrm{P}}(\eta)n\tau/2}\rrangle$, which gives rise to an asymptotic stretched-exponential behavior arising from the inhomogeneously broadened Purcell decay:
\begin{equation}
    \llangle e^{-\frac{\Gamma_{\mathrm{P}(\eta)}}{2}n\tau}\rrangle=\int_{-\infty}^\infty d\eta\:\frac{T_2^*}{\sqrt{4\pi}}e^{-\frac{1}{4}\eta^2(T_2^*)^2}e^{-\frac{\Gamma_{\mathrm{P}}(\eta)}{2}n\tau}\sim e^{\left(\frac{\kappa T_2^*}{4}\right)^2}e^{-\sqrt{\gamma_{\mathrm{P}}n\tau}},\quad n\tau\to\infty,\label{eq-supp-mat:gammap}
\end{equation}
where
\begin{equation}
    \gamma_{\mathrm{P}}\simeq (gT_2^*)^2\frac{\kappa}{2},\quad T_2^*\lvert\delta\rvert\ll1.\label{eq-supp-mat:gammap2}
\end{equation}
The results displayed in Eqs.~\eqref{eq-supp-mat:gammap} and \eqref{eq-supp-mat:gammap2} were obtained by noting that the integrand ($\propto \exp[-F(\eta)]$) approaches the sum of two Gaussians centered at $\eta=\pm\eta_0$ in the limit $n\tau\to\infty$, where $\eta_0$ is a stationary point [$F'(\eta_0)=0$].  

Combining Eqs.~\eqref{eq-supp-mat:sigma-minus-2} and \eqref{eq-supp-mat:gammap}, the total echo envelope is thus given at any echo $t=n\tau$ (for $\kappa T_2^*<1$) by
\begin{equation}
    \tilde{C}(n\tau)=\braket{\tilde{\sigma}_-}_{n\tau}/\braket{\tilde{\sigma}_-}_0\simeq e^{-\sqrt{\gamma_{\mathrm{P}}n\tau}}\tilde{C}_0(n\tau),\quad (\kappa T_2^*<1).
\end{equation}
If $\tilde{C}_0(t)$ is slowly-varying on the timescale $\sim T_2^*$ of each echo, this gives:
\begin{equation}
    \braket{\tilde{\sigma}_-}_t\simeq\braket{\sigma_-}_0\sum_n G_{n}(t-n\tau)\tilde{C}(n\tau),\label{eq-supp-mat:coherence-expansion}
\end{equation}
where 
\begin{equation}
    \boxed{G_n(t) = e^{\sqrt{\gamma_{\mathrm{P}}n\tau}}\left\llangle e^{-\Gamma_\mathrm{P}(\eta)n\tau/2}e^{-i\eta t}\right\rrangle.}\label{eq-supp-mat:revival-shape}
\end{equation}
The same asymptotic analysis described above can be used to find the shape of $G_n(t)$ at short and long times (for $\kappa T_2^*<1$): 
\begin{equation}
    G_{n}(t)\simeq\begin{cases}
    e^{-\left(\frac{t}{T_2^*}\right)^2},\quad &\gamma_{\mathrm{P}}n\tau<1\\
    e^{-\left(\frac{t}{2T_2^*}\right)^2}\cos{\left[\sqrt{2}(\gamma_{\mathrm{P}}n\tau)^{1/4}\frac{t}{T_2^*}\right]},\quad &\gamma_{\mathrm{P}}n\tau>1
    \end{cases}.\label{eq-supp-mat:revival-profile}
\end{equation}
Remarkably, the shape of the echoes changes as $n$ increases: For $n\tau<1/\gamma_\mathrm{P}$, the echoes are simply Gaussians broadened by $T_2^*$, but for $n\tau>1/\gamma_\mathrm{P}$, the width of the echo peaks doubles, and moreover, the peaks show a cosine modulation due to a combination of Purcell decay and inhomogeneous broadening [Fig.~\ref{fig:asymptotics}].

\begin{figure}
    \centering
    \includegraphics[width=0.8\textwidth]{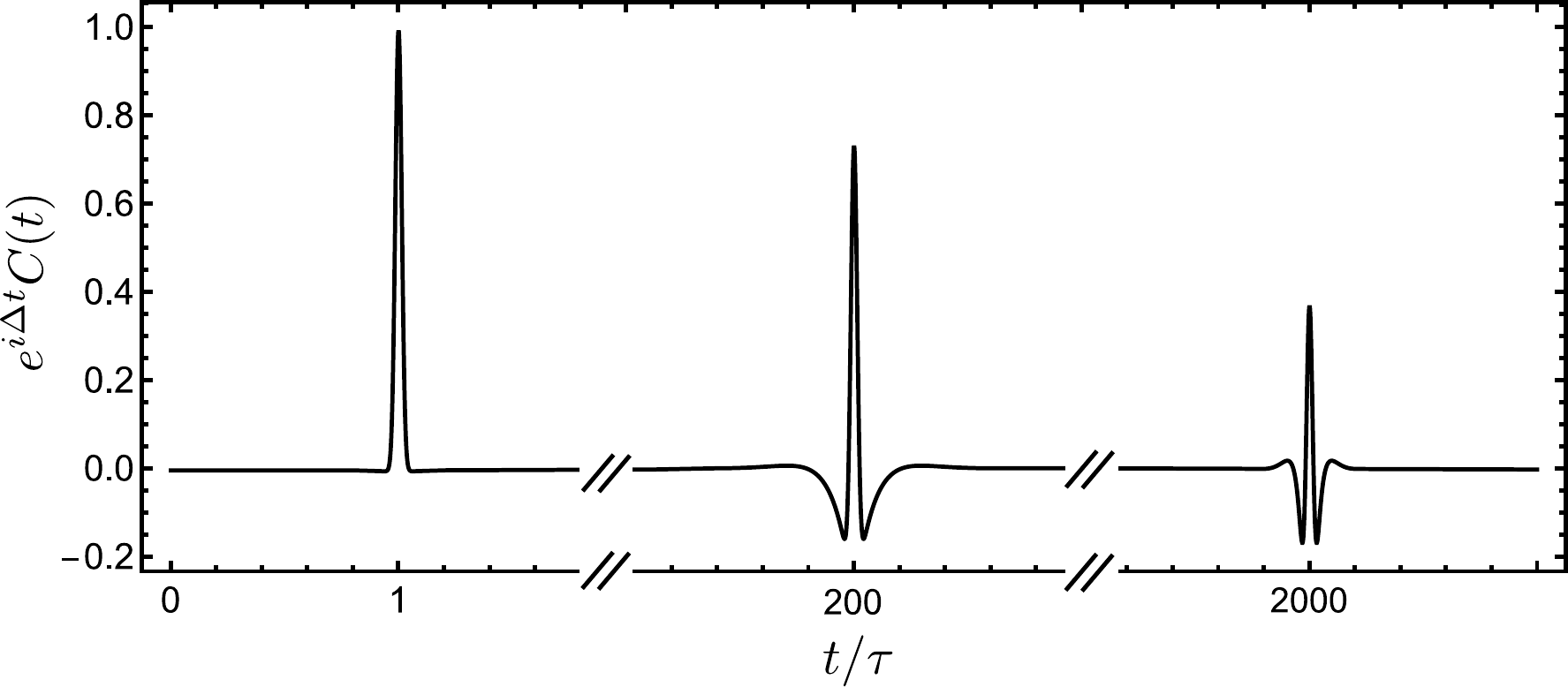}\hspace{1.1cm}
    \caption{Solid line: Revivals described by $e^{i\Delta t}C(t)=e^{i\Delta t}\braket{\sigma_-}_t/\braket{\sigma_-}_0$ [evaluated from Eq.~\eqref{eq-supp-mat:coherence-expansion} by numerically integrating Eq.~\eqref{eq-supp-mat:revival-shape} for $G_n(t)$, assuming $\Delta\tau=2\pi m$ for some $m\in\mathbb{Z}$ so that $e^{i\Delta t}C(t)$ is real] for three different values of $n$: $n=1, 200, 2000$. We assume as in Fig.~\ref{fig:backaction} that $g=0.1\kappa$, $\kappa T_2^*=0.1$, and $\kappa\tau=10$, giving $\gamma_{\mathrm{P}}^{-1}=2000\tau$. For $\gamma_{\mathrm{P}}n\tau\gtrsim1$, the revivals are suppressed by a stretched exponential. In addition, the Gaussian envelope broadens and becomes modulated by a cosine due to an interplay of inhomogeneous broadening and Purcell decay.}
    \label{fig:asymptotics}
\end{figure}

Recalling the relation between the lab-frame expectation value $\braket{\sigma_-}_t$ and the toggling-frame observables $\braket{\tilde{\sigma}_\pm}_t$ [Eq.~\eqref{eq-supp-mat:lab-toggle}], we use Eq.~\eqref{eq-supp-mat:coherence-expansion} to determine that 
\begin{equation}
    e^{i\Delta t}\braket{\sigma_-}_t\simeq \frac{1}{2}\braket{\sigma_x}_0\sum_{n=0}^N G_n(t-n\tau)e^{i\Delta n\tau}\begin{cases}
        \tilde{C}(n\tau),\quad &n\text{ even}\\
        \tilde{C}^*(n\tau),\quad &n\text{ odd}
    \end{cases},\label{eq-supp-mat:sigma-expectation-time}
\end{equation}
where we have used the fact that for an initial $(-\pi/2)_y$ pulse on the qubit, $\braket{\sigma_-}_0=\frac{1}{2}\braket{\sigma_x}_0$. Substituting Eq.~\eqref{eq-supp-mat:sigma-expectation-time} into Eq.~\eqref{eq-supp-mat:cavity-output-high-Q} gives
\begin{equation}
    \braket{\tilde{a}}_t\simeq\frac{1}{2}\braket{\sigma_x}_0\left[\frac{1}{2}f_0(t)+\sum_{n=1}^Nf_n(t-n\tau)e^{i\Delta n\tau}\mathcal{K}^n\tilde{C}(n\tau)\right],\quad \kappa\tau\gg 1,\label{eq-supp-mat:train-of-sharks}
\end{equation}
where we have introduced the complex conjugation operator $\mathcal{K}$ ($\mathcal{K}z=z^*, z\in\mathbb{C}$), and where the wavepackets $f_n(t)$ are given by
\begin{equation}\label{eq-supp-mat:sharkfin}
    f_n(t)=-ig\int_{-\infty}^tdt' \chi_c(t-t')G_n(t').
\end{equation}
The first term, $(1/2)f_0(t)$, in Eq.~\eqref{eq-supp-mat:train-of-sharks} reflects the fact that revivals with $n\geq 1$ have twice as much ``area'' as the initial free-induction decay. For $\kappa T_2^*\ll1$, we can replace $G_n(t)$ in Eq.~\eqref{eq-supp-mat:sharkfin} by a delta function with appropriate weight: 
\begin{equation}
    G_n(t)\simeq\delta(t)\sqrt{\pi}T_2^*\bar{G}_n;\quad \bar{G}_n=\frac{1}{\sqrt{\pi}T_2^*}\int_{-\infty}^\infty dt \:G_n(t). 
\end{equation}
In this case, we have the simplified form
\begin{equation}\label{eq-supp-mat:sharkfin-simplified}
    f_n(t)\simeq-i\sqrt{\pi}gT_2^*\bar{G}_n\chi_c(t)=-i\sqrt{\pi}gT_2^*\bar{G}_ne^{-i\delta t-\frac{\kappa}{2}t}\Theta(t), \quad \kappa T_2^*\ll 1.
\end{equation}
The amplitude of the $n^\mathrm{th}$ revival in the cavity field due to a CPMG sequence is thus suppressed by the dimensionless factor $\bar{G}_n$ \emph{in addition} to any suppression due to coherence decay. From the asymptotic forms given in Eq.~\eqref{eq-supp-mat:revival-profile}, we have
\begin{equation}
    \bar{G}_n\simeq\begin{cases}1,& n\ll 1/(\gamma_\mathrm{P}\tau)\\
    2 e^{-2\sqrt{\gamma_\mathrm{P}n\tau}},& n\gg 1/(\gamma_\mathrm{P}\tau)
    \end{cases}.
\end{equation}
Even in the absence of external sources of dephasing [$\gamma_\phi=h(t)=0$], Purcell decay in $\tilde{C}(n\tau)\simeq e^{-\sqrt{\gamma_\mathrm{P}n\tau}}$ together with the additional suppression of $\bar{G}_n$ due to the cosine modulation of echoes [Eq.~\eqref{eq-supp-mat:revival-profile}] results in an asymptotic suppression of cavity revivals:
\begin{equation}\label{eq-supp-mat:GnCAsymptotic}
    \bar{G}_n\tilde{C}(n\tau)\sim 2e^{-3\sqrt{\gamma_\mathrm{P}n\tau}},\quad \left[\gamma_\phi=h(t)=0, n\tau\gg 1/\gamma_\mathrm{P}\right].
\end{equation}

Taking the Fourier transform of Eq.~\eqref{eq-supp-mat:train-of-sharks} and applying the approximations given above for $\kappa T_2^*\ll 1$ gives
\begin{equation}
    \braket{\tilde{a}}_{\omega}\simeq-i\frac{\braket{\sigma_x}_0}{2}\sqrt{\pi}gT_2^*\chi_c(\omega)\left[\frac{1}{2}+\sum_{n=1}^N e^{i(\omega+\Delta)n\tau}\bar{G}_n\mathcal{K}^n \tilde{C}(n\tau)\right],\quad T_2^*\kappa\ll1, \label{eq-supp-mat:cavity-field-CPMG}
\end{equation}
where $\chi_c(\omega)=[i(\delta-\omega)+\kappa/2]^{-1}$. Since $\chi_c(\omega)$ is peaked at $\omega=\delta$, we can maximize the signal by considering $\omega=\delta$. This gives Eq.~\eqref{eq:spectrum-peak} of the main text,
\begin{equation}
    \boxed{\braket{\tilde{a}}_{\omega=\delta}\simeq-i\braket{\sigma_x}_0\frac{\sqrt{\pi}gT_2^*}{\kappa}\left[\tilde{C}_{N,\tau}(\delta)-\frac{1}{2}\right],}\label{eq-supp-mat:cavity-spectrum-at-delta}
\end{equation}
where 
\begin{equation}
    \boxed{\tilde{C}_{N,\tau}(\omega)=\sum_{n=0}^N e^{in\omega\tau}[e^{in\delta_\Delta\tau}\bar{G}_n\mathcal{K}^n\tilde{C}(n\tau)] }\label{eq-supp-mat:dft}
\end{equation} 
is the discrete Fourier transform of $e^{in\delta_\Delta\tau}\bar{G}_n\mathcal{K}^n\tilde{C}(n\tau)$. We have written Eq.~\eqref{eq-supp-mat:dft} in terms of $\delta_\Delta$, where $\delta_\Delta\equiv\Delta$ (mod $2\pi/\tau$), in order to emphasize that, due to the periodicity of $e^{i\Theta}$ ($\Theta\in\mathbb{R}$), the shift by $\Delta$ of the frequency content of the echo envelope is equivalent to a phase shift due to a smaller quantity $\delta_\Delta$ bounded above by $2\pi/\tau$. [In the case where $\tau=2\pi m/\Delta$ for some $m\in\mathbb{Z}$, $\delta_\Delta=0$.] Provided both $\Delta$ and $\tau$ are known, the revival amplitudes $\bar{G}_n\mathcal{K}^n\tilde{C}(n\tau)$ can be recovered by sweeping the detuning over some interval $O(2\pi/\tau)$, inverting the discrete Fourier transform, multiplying by $\bar{G}_n^{-1}$, and, in the case where $\tau\neq 2\pi m/\Delta$, multiplying by the appropriate phase factor ($e^{-i\delta_\Delta n\tau}$).

\section{Quantum noise}
\label{supp-mat-sec:S4-quantum-noise}

We can find a generic expression for the echo envelope $\tilde{C}_0(n\tau)$ [Eq.~\eqref{eq-supp-mat:C0}] within a leading-order Magnus expansion and Gaussian approximation. These approximations are generic and conditions for their validity will be satisfied for a broad class of non-Markovian environments. 

At $t=0$, the qubit is subjected to a fast $(-\pi/2)_y$-pulse followed by a dynamical decoupling sequence (starting at $t=0^+$) consisting of fast $\pi_x$ pulses. As shown in Sec.~\ref{supp-mat-sec:S3-qubit-drive} [Eq.~\eqref{eq-supp-mat:C0}], the echo envelope is described by
\begin{equation}
    \tilde{C}_0(t)=e^{-\gamma_\phi t}\braket{\tilde{U}_-^\dagger (t)\tilde{U}_+(t)},\label{eq-supp-mat:C0-also}
\end{equation}
where [recalling Eq.~\eqref{eq-supp-mat:conditional-unitaries}] $\tilde{U}_\pm(t)=\mathcal{T}e^{-i\int_0^t dt'\tilde{V}_\pm(t')}$ for $\tilde{V}_\pm(t)=[h(t)\pm\delta\Omega(t)s(t)]/2$. Provided the Magnus expansion converges, the unitaries $\tilde{U}_\pm(t)$ can in turn be expressed using the Magnus expansion as
\begin{equation}
    \tilde{U}_\pm(t)=e^{-i\Lambda_\pm(t)},\quad \Lambda_\pm(t)=\sum_{k=1}^\infty \Lambda_\pm^{(k)}(t),\label{eq-supp-mat:magnus}
\end{equation}
where, in terms of $h(t)=e^{i(H_{\mathrm{E}}-h/2)t}he^{-i(H_{\mathrm{E}}-h/2)t}$, the first two terms in the sum are given by
\begin{align}
    &\Lambda_\pm^{(1)}(t)=\int_0^t dt'\;\frac{1}{2}[(1\pm s(t'))h(t')\pm  \delta\eta(t')s(t')],\\
    &\Lambda_\pm^{(2)}(t)=-\frac{i}{8}\int_0^t dt'\int_0^{t'}dt''\:(1\pm s(t'))(1\pm s(t''))[h(t'),h(t'')].
\end{align}
[A sufficient condition for convergence of the Magnus expansion is $\int_0^t dt'\lvert \tilde{V}_\pm(t')\rvert <\pi$~\cite{S-blanes2009magnus}, which can be satisfied for weak coupling to an environment, e.g.~spins or charge fluctuators, having a bounded spectrum.] Substituting Eq.~\eqref{eq-supp-mat:magnus} into Eq.~\eqref{eq-supp-mat:C0-also}, we then find using the Baker-Campbell-Hausdorff  formula that, neglecting $O(h^3)$ and higher,
\begin{equation}
    \tilde{C}_0(t)\simeq e^{-\gamma_\phi t}\braket{e^{-i\Lambda_{\mathrm{eff}}(t)}},
\end{equation}
where
\begin{align}
    \Lambda_{\mathrm{eff}}(t)=\Lambda_+(t)-\Lambda_-(t)+\frac{1}{2i}[\Lambda_+(t),\Lambda_-(t)].
\end{align}
$\Lambda_{\mathrm{eff}}(t)$ is Hermitian, which can easily be verified using the fact that $\Lambda_\pm^\dagger(t)=\Lambda_\pm(t)$.
Making a Gaussian approximation, i.e.~neglecting third- and higher-order cumulants in $h(t)$~\cite{S-beaudoin2013enhanced}, and, without loss of generality, taking $\braket{\Lambda_\pm^{(1)}(t)}=0$ gives
\begin{equation}
    \boxed{\tilde{C}_0(t)\simeq e^{-\gamma_\phi t}e^{-i\Phi_{\mathrm{q}}(t)-\chi(t)},}\label{eq-supp-mat:c0-3}
\end{equation}
where, in terms of $\delta \Omega(t)=\delta \eta(t)+h(t)$, the reduction in amplitude is described by
\begin{align}
    \chi(t)&=\frac{1}{2}\braket{[\Lambda_+^{(1)}(t)-\Lambda_-^{(1)}(t)]^2}\\
    &=\frac{1}{4}\int_0^t dt'\int_0^t dt'' s(t')s(t'')\braket{\{\delta \Omega(t'),\delta \Omega(t'')\}}\\
    &=\frac{1}{2}\int_0^t dt'\int_0^t dt'' s(t')s(t'')\mathrm{Re}\braket{\delta\Omega(t')\delta\Omega(t'')}\\
    &=\frac{1}{2}\int_0^t dt'\int_0^t dt'' s(t')s(t'')\mathrm{Re}\braket{\delta\Omega(\lvert t'-t''\rvert)\delta\Omega}.\label{eq-supp-mat:chi}
\end{align}
In the last line above, we used stationarity of the correlation function $\braket{\delta\Omega(t')\delta\Omega(t'')}$ for our choice of initial state, $[H_{\mathrm{E}}-h/2,\bar{\rho}_{\mathrm{E}}]=0$, as well as symmetry of $\mathrm{Re}\braket{\delta\Omega(t')\delta\Omega(t'')}$ under interchange of $t'\leftrightarrow t''$, so that $\mathrm{Re}\braket{\delta\Omega(t')\delta\Omega(t'')}=\mathrm{Re}\braket{\delta\Omega(\lvert t'-t''\rvert)\delta\Omega}$. For CPMG and $t=n\tau$, this result for $\chi$ can equivalently be written in terms of $\Omega(t)=\eta+\delta\Omega(t)$ [by replacing $\delta\Omega$ with $\Omega$ in Eq.~\eqref{eq-supp-mat:chi}] since $\int_0^{n\tau} dt s(t)\eta=0$.

In addition to the reduction in amplitude, there is also a phase given by
\begin{equation}
    \Phi_{\mathrm{q}}(t)=\braket{\Lambda_+^{(2)}(t)-\Lambda_-^{(2)}(t)}+\frac{1}{2i}\braket{[\Lambda_+^{(1)}(t),\Lambda_-^{(1)}(t)]}.
\end{equation}
Evaluating the two terms separately gives
\begin{align}
    \braket{\Lambda_+^{(2)}(t)-\Lambda_-^{(2)}(t)}=\frac{1}{2}\int_0^t dt'\int_0^{t'}dt''[s(t')+s(t'')]\mathrm{Im}\braket{h(t'-t'')h}\label{eq-supp-mat:quantum-phase-1}
\end{align}
and
\begin{align}
    \frac{1}{2i}\braket{[\Lambda_+^{(1)}(t),\Lambda_-^{(1)}(t)]}=\frac{1}{2}\int_0^t dt'\int_0^{t'}dt''[s(t')-s(t'')]\mathrm{Im}\braket{h(t'-t'')h},\label{eq-supp-mat:quantum-phase-2}
\end{align}
where we have replaced $(2i)^{-1}\braket{[h(t'),h(t'')]}=\mathrm{Im}\braket{h(t')h(t'')}=\mathrm{Im}\braket{h(t'-t'')h}$. Adding Eqs.~\eqref{eq-supp-mat:quantum-phase-1} and \eqref{eq-supp-mat:quantum-phase-2} then gives the total phase,
\begin{equation}
    \Phi_{\mathrm{q}}(t)=\int_0^t dt'\int_0^{t'}dt'' s(t')\mathrm{Im}\braket{h(\lvert t'-t''\rvert)h}=\int_0^t dt'\int_0^{t'}dt''s(t')\mathrm{Im}\braket{\delta\Omega(\lvert t'-t''\rvert)\delta\Omega},\label{eq-supp-mat:quantum-phase}
\end{equation}
where introducing the absolute value in $\delta\Omega(\lvert t'-t''\rvert)$ has no effect on the integral since $t'\geq t''$ over the entire region of integration. [It does however allow us to write the classical and quantum parts in a symmetric way in Eq.~\eqref{eq-supp-mat:spectral-density}, below.] Given that $\Phi_{\mathrm
q}(t)$ depends on $\mathrm{Im}\braket{\delta\Omega(\lvert t\rvert)\delta\Omega}=\braket{[\delta\Omega(\lvert t\rvert),\delta\Omega]}/(2i)$, it can \textit{only} arise from a quantum environment. This phase was also derived in Refs.~\cite{S-kwiatkowski2020influence, S-pazsilva2017multiqubit}, but for the case where $[H_{\mathrm{E}},\bar{\rho}_{\mathrm{E}}]=0$ and for coupling  of the form $h\ket{e}\bra{e}$. $\Phi_{\mathrm{q}}$ was however found to vanish identically for the case where $[H_{\mathrm{E}},\bar{\rho}_{\mathrm{E}}]=0$ and for coupling $h\sigma_z/2$~\cite{S-kwiatkowski2020influence, S-pazsilva2017multiqubit}, highlighting the importance of the initial state (here arising from preparation of $\bar{\rho}_{\mathrm{E}}$ with the qubit in $\ket{g}$) in determining whether or not a particular form of the coupling leads to a dependence on quantum noise. 

We now introduce the spectral density
\begin{equation}
    \boxed{S(\omega)=S_{\mathrm{c}}(\omega)+iS_{\mathrm{q}}(\omega)=\mathrm{lim}_{\epsilon\rightarrow 0^+}\int_{-\infty}^{\infty}dt\:e^{-i\omega t-\epsilon\lvert t\rvert}\braket{\Omega(\lvert t\rvert)\Omega},}\label{eq-supp-mat:spectral-density}
\end{equation}
where the infinitesimal parameter $\epsilon$ ensures convergence of the integral. Equation \eqref{eq-supp-mat:spectral-density} corresponds to Eq.~\eqref{eq:spectral-density} of the main text. In general, there are two contributions to $S(\omega)$: a classical part $S_{\mathrm{c}}(\omega)=\mathrm{Re}\:S(\omega)=S_h(\omega)+S_\eta(\omega)$ that sets the magnitude ($\sim e^{-\chi}$) of $\tilde{C}_0(t)$, and a quantum part $S_{\mathrm{q}}(\omega)=\mathrm{Im}\:S(\omega)$ that contributes a phase. The classical and quantum parts $S_h(\omega)$ and $S_{\mathrm{q}}(\omega)$ depend, respectively, on symmetrized and anti-symmetrized correlation functions:
\begin{align}
    S_h(\omega)&=\frac{1}{2}\int \frac{dt}{2\pi} e^{-i\omega t}\braket{\{h(\lvert t\vert),h\}},\\
    S_{\mathrm{q}}(\omega)&=\mathrm{lim}_{\epsilon\rightarrow 0^+}\frac{1}{2i}\int \frac{dt}{2\pi} e^{-i\omega t-\epsilon \lvert t\rvert}\braket{[h(\lvert t\rvert),h]}.
\end{align}
For $\int_0^t dt' s(t')=0$ (as would occur for CPMG and $t=n\tau$), the quantities $\chi(t)$ and $\Phi_{\mathrm{q}}(t)$ appearing in $\tilde{C}_0(t)$ [Eq.~\eqref{eq-supp-mat:c0-3}] can be written in terms of $S_{\mathrm{c,q}}(t)$ as
\begin{equation}
    \boxed{\chi(t)=\int \frac{d\omega}{2\pi}\frac{F_{\mathrm{c}}(\omega, t)}{\omega^2}S_{\mathrm{c}}(\omega)}
\end{equation}
and
\begin{equation}
    \boxed{\Phi_{\mathrm{q}}(t)=\int\frac{d\omega}{2\pi}\frac{F_{\mathrm{q}}(\omega,t)}{\omega^2}S_{\mathrm{q}}(\omega),}
    \label{eq-supp-mat:echo-env}
\end{equation}
corresponding to Eqs.~\eqref{eq:phi} and \eqref{eq:chi} of the main text. Here, we have introduced classical-noise (c) and quantum-noise (q) filter functions  given by
\begin{equation}
F_{\mathrm{c}}(\omega,t) = \frac{\omega^2}{2}\left|\int_0^{t}dt'\:e^{i\omega t'}s(t')\right|^2
\end{equation}
and 
\begin{equation}
    F_{\mathrm{q}}(\omega,t)=\omega \int_0^t dt'\:\mathrm{sin}(\omega t')s(t').
\end{equation}

In noise spectroscopy, it is commonly assumed that $S_{\mathrm{q}}(\omega)=0$ and consequently, that $\tilde{C}_0(n\tau)$ is real-valued and positive~\cite{S-szankowski2017environmental, S-alvarez2011measuring, S-yuge2011measurement, S-bylander2011noise, S-szankowski2019transition}. However, ignoring the quantum-noise correction when it is significant can lead to erroneous conclusions. In these cases, the simple generalization provided by Eq.~\eqref{eq-supp-mat:echo-env} can be used to perform accurate noise spectroscopy, even when the quantum noise is significant.

\section{Modeling the effect of inhomogeneous broadening on the cavity transmission spectrum}
\label{supp-mat-sec:s5-input-output}

\noindent In order to calculate the cavity transmission spectrum $A_\mathrm{T}(\omega)=r_{\mathrm{out},2}(\omega)/r_{\mathrm{in},1}(\omega)$ using input-output theory~\cite{S-gardiner1985input}, one must consider the coupled quantum Langevin equations of the cavity and (qubit-environment) system. We assume no qubit driving, $H_{\mathrm{drive}}(t)=0$. Within a rotating-wave approximation (RWA) requiring that $g$, $\lvert\delta\rvert\ll\lvert\Delta+\omega_c\rvert$, these equations can be decoupled in linear-response with respect to $g$~\cite{S-kohler2018dispersive, S-mielke2021nuclear}, resulting in a qubit susceptibility [Eq.~\eqref{eq-supp-mat:input-output-susceptibility}, below] that depends only on the eigenstates $\ket{\sigma,m}$ of $H_{\sigma}=\bra{\sigma}\frac{1}{2}h\sigma_z\ket{\sigma}+H_{\mathrm{E}}$ for a fixed value of $\sigma\in\{e,g\}$. The effect of inhomogeneous broadening can then be modeled by averaging $A_\mathrm{T}(\omega)=\llangle A_\mathrm{T}(\omega,\eta)\rrangle$ over the distribution $\llangle\dotsm\rrangle$ of noise realizations. For a time-independent $\eta$, this gives
\begin{equation}
    A_{\mathrm{T}}(\omega)\simeq\bigg\langle\hspace{-0.15cm}\bigg\langle\frac{-\sqrt{\kappa_1\kappa_2}}{i(\omega_c-\omega)+ig^2\chi_\eta(\omega)+\kappa/2}\bigg\rangle\hspace{-0.15cm}{\bigg\rangle}, \label{eq-supp-mat:input-output-transmission}
\end{equation} 
where
\begin{equation}
    \chi_\eta(\omega)=i\sum_{mn}\frac{(p_{en}-p_{gm})\lvert\braket{g,m\rvert e,n}\rvert^2}{i(\Delta-\omega+\eta-(\epsilon_{gm}-\epsilon_{en}))+\gamma_\phi}.\label{eq-supp-mat:input-output-susceptibility}
\end{equation}
In Eq$.$~(\ref{eq-supp-mat:input-output-susceptibility}), we denote by $\epsilon_{\sigma m}$ the eigenenergies of the Hamiltonian $H_{\sigma}$: $H_{\sigma}\ket{\sigma, m}=\epsilon_{\sigma m}\ket{\sigma, m}$. This result also assumes an initial state $\rho(0)$ that is diagonal in the eigenbasis of $\sum_{\sigma}H_{\sigma}\ket{\sigma}\bra{\sigma}$, so that $\rho(0)=\sum_{\sigma,m} p_{\sigma m}\ket{\sigma}\bra{\sigma}\otimes\ket{\sigma, m}\bra{\sigma, m}$.

\section{Quantifying the signal}
\label{supp-mat-sec:s6-signal}

After each measurement cycle, typically involving an $N$-pulse dynamical decoupling sequence, the transmission line coupled to the output port will be in a quantum state  $\rho_{\mathrm{TL}}$ that encodes information about the coherence dynamics that occurred throughout the dynamical decoupling sequence. The amount of information that can be gained per measurement cycle will be limited by both the nature of the (generally mixed) state $\rho_{\mathrm{TL}}$ and by the inference procedure used to extract information from it. Here, we characterize a measure of the signal that depends only on $\rho_{\mathrm{TL}}$. 

Provided $\rho_\mathrm{TL}$ can be described in the subspace of zero or one photons (this limit can always be reached by taking $\kappa_2/\kappa$ sufficiently small), we can write it in the general form
\begin{equation}
    \boxed{\rho_{\mathrm{TL}}=(1-S)\rho_{\mathrm{inc}}+S\ket{\Psi}\bra{\Psi},}\label{eq-supp-mat:trans-line-dm-2}
\end{equation}
where the incoherent part $\rho_\mathrm{inc}$ satisfies $\mathrm{Tr}\{r_{k,2}\rho_{\mathrm{inc}}\}=0\;\forall k$. In Eq.~\eqref{eq-supp-mat:trans-line-dm-2}, the size of the signal is characterized by $S\in[0,1]$, and the coherence is fully described by the state $\ket{\Psi}$ of an effective two-level system:
\begin{equation}
    \ket{\Psi}=\frac{1}{\sqrt{2}}\left(\ket{0}+\ket{1}\right);\quad \ket{1}=\frac{2}{S}\sum_k \braket{r_{k,2}}_t r_{k,2}^\dagger\ket{0};\quad S=2[\sum_k\lvert \braket{r_{k,2}}_t\rvert^2]^{1/2}.\label{eq-supp-mat:signal}
\end{equation}

To interpret the meaning of the signal $S$, it is useful to consider an extreme example. We consider a qubit prepared in an initial state determined by some fixed (but initially undetermined) phase $\phi_0$: $\braket{\sigma_-}_0=e^{-i\phi_0}/2$. The qubit is then coupled to the cavity, but otherwise has no source of dephasing: $\gamma_\phi=h(t)=\eta(t)=0$. We assume for this example that there is no additional dynamics induced through dynamical decoupling [$H_\mathrm{drive}(t)=0$] and furthermore take $\kappa_1=\kappa_\mathrm{ext}=0$. After a time $t\gg \Gamma_{\mathrm{P}}^{-1}$, the state of the qubit will have been transferred, via a Wigner-Weisskopf decay process, to a definite pure state $\ket{\Psi}=\ket{\Psi(\phi_0)}$ of the output transmission line,  giving $S=1$. The initial phase $\phi_0$ of the qubit can then be inferred (in principle) through a phase estimation procedure by performing measurements on a well defined two-level subspace of transmission-line states, yielding up to one bit per measurement. In general, the state of the transmission line may be correlated with the state of the qubit, environment, and cavity. These correlations, together with the average $\llangle \rrangle$ over realizations of the random noise parameter $\eta(t)$, will lead to a mixed state $\rho_\mathrm{TL}$ with $S<1$. Having a reduced value $S<1$ thus sets a fundamental limitation on the information that can be extracted from the complete state of the transmission line. It is straightforward to characterize the maximum achievable signal $S$ given the coefficients $\braket{r_{k,2}}_t$ arising from a CPMG sequence, fully accounting for correlations with other degrees of freedom and accounting for random noise. We now proceed with this task.

Limits on the signal $S$ can be found in the present context from the sequence of derivations given above. We substitute Eq.~\eqref{eq-supp-mat:sharkfin-simplified} for the approximate form of the wavepackets $f_n(t)$ (for $\kappa T_2^*\ll 1$) into Eq.~\eqref{eq-supp-mat:train-of-sharks} for the cavity field $\braket{\tilde{a}}_t$. This result is then substituted into Eq.~\eqref{eq-supp-mat:eom-rk} for $\braket{r_{k,2}}_t$, which, for $\braket{r_{k,2}}_0=0$, gives
\begin{equation}\label{eq-supp-mat:rk}
    \braket{r_{k,2}}_t\simeq -\frac{1}{2}\braket{\sigma_x}_0\sqrt{\pi}g T_2^*\eta_{k,2}\left[\frac{1}{2}X_{0k}(t)+\sum_{n=1}^N \bar{G}_nX_{nk}(t)\mathcal{K}^n\tilde{C}(n\tau)\right], 
\end{equation}
where (recalling the relation $\delta = \omega_c-\Delta$),
\begin{equation}
    X_{nk}(t)=e^{i\omega_k(t-n\tau)}\int_0^tdt'e^{-i(\omega_k-\omega_c)t'-\frac{\kappa}{2}t'}\Theta(t'-n\tau).
\end{equation}
For times $t-n\tau\gg\kappa^{-1}$ long compared to the timescale over which cavity transients die out, this object takes the simple form
\begin{equation}
    X_{nk}(t)\simeq \frac{e^{-i\omega_k(t-n\tau)}}{\frac{\kappa}{2}-i(\omega_k-\omega_c)}\Theta(t-n\tau);\quad t-n\tau\gg \kappa^{-1}.
\end{equation}
To evaluate $S$, we substitute the expression for $\braket{r_{k,2}}_t$ [Eq.~\eqref{eq-supp-mat:rk}] into $\sum_k \left|\braket{r_{k,2}}_t\right|^2$. In addition to terms arising from the same echo/revival, proportional to
\begin{equation}
    \sum_k |\eta_{k,2}|^2\left|X_{nk}(t)\right|^2\simeq \sum_k \frac{|\eta_{k,2}|^2}{(\omega_k-\omega_c)^2+(\kappa/2)^2}\Theta(t-n\tau)\simeq\frac{\kappa_2}{\kappa},\quad t-n\tau\gg \kappa^{-1},
\end{equation}
there will also be cross terms associated with distinct echoes at times $n\tau$, $m\tau$, with $n\ne m$. These cross-terms will, however, be suppressed exponentially for $\kappa\tau\gg 1$:
\begin{equation}
    \sum_k |\eta_{k,2}|^2X_{nk}(t)X_{mk}^*(t)\simeq \frac{\kappa_2}{\kappa} e^{-|n-m|\kappa\tau/2}\Theta(t-n\tau)\Theta(t-m\tau)\simeq 0;  \quad (n\ne m, \kappa\tau\gg 1).
\end{equation}
Neglecting these cross terms for $\kappa\tau\gg 1$, we then find that
\begin{equation}
    S=\left[\left|\braket{\sigma_x}_0\right|^2\pi(gT_2^*)^2 \frac{\kappa_2}{\kappa} N_\mathrm{eff}\right]^{1/2},
\end{equation}
where the parameter $N_\mathrm{eff}$ scales with the number of revivals/echoes that can be achieved before coherence is lost:
\begin{equation}
N_\mathrm{eff} = \frac{1}{4}+\sum_{n=1}^N \left|\bar{G}_n\tilde{C}(n\tau)\right|^2,\quad t-N\tau>\kappa^{-1}.
\end{equation}

For a Hahn echo sequence ($N=1$), the maximum signal is achieved for $|\braket{\sigma_x}_0|=1$ and $|\bar{G}_1\tilde{C}(\tau)|=1$, giving
\begin{equation}
   \boxed{ S \le S_\mathrm{Hahn} = \frac{\sqrt{5\pi}}{2} gT_2^*\sqrt{\frac{\kappa_2}{\kappa}}.}
\end{equation}
In this case, the total recoverable signal $S$ per cycle is thus limited by $gT_2^*\ll 1$. In the case of a CPMG sequence, we expect the product $|\bar{G}_n\tilde{C}(n\tau)|$ to be upper-bounded in the best case by the asymptotic form given in Eq.~\eqref{eq-supp-mat:GnCAsymptotic}, resulting in
\begin{eqnarray}
    N_\mathrm{eff} & \le & \frac{1}{4}+\sum_{n=1}^\infty 4e^{-6\sqrt{\gamma_\mathrm{P}n\tau}}\\
    & \simeq & \frac{4}{\gamma_\mathrm{P}\tau}\int_0^\infty dx e^{-6\sqrt{x}}= \frac{2}{9\gamma_\mathrm{P}\tau},\quad (\gamma_\mathrm{P}\tau\ll 1).
\end{eqnarray}
Inserting this result into the definition for $S$ and using the relation $\gamma_\mathrm{P}=(gT_2^*)^2\kappa/2$ gives an approximate upper bound on the signal that can be achieved with a CPMG sequence:
\begin{equation}
    \boxed{S\lesssim S_\mathrm{CPMG} = \frac{2\sqrt{\pi}}{3}\sqrt{\frac{\kappa_2}{\kappa}\frac{1}{\kappa\tau}}.}\label{eq-supp-mat:SCPMG}
\end{equation}
For the CPMG sequence, the recoverable signal is not limited by $gT_2^*\ll1$, but it is still small in the parameter $1/\kappa\tau\ll 1$.

We can improve on the result given in Eq.~\eqref{eq-supp-mat:SCPMG} by modulating the coupling $g\rightarrow g(t)$ [or the detuning $\delta\rightarrow \delta(t)$] as a function of time so that $g(t)\neq 0$ [$\delta(t)\lesssim (T_2^*)^{-1}$] only for times $\lvert t-n\tau\rvert\leq t_{\mathrm{on}}$, where $t_{\mathrm{on}}<T_2^*$ is short compared to the duration of a revivial. This has the effect of reducing cavity-induced backaction on the qubit by eliminating incoherent Purcell decay at a rate $\Gamma_{\mathrm{P}}=g^2\kappa/[(\eta-\delta)^2+(\kappa/2)^2]$ for times when the qubit coherence is already suppressed by inhomogeneous broadening. In this case, the self-energy [Eq.~\eqref{eq-supp-mat:self-energy}] is given by
\begin{equation}
    \Sigma(t,t')\simeq -ig^2\sum_n\Theta_n(t)\Theta_n(t'),
\end{equation}
where $\Theta_n(t)=\Theta\left(t-\left(n\tau-\tfrac{t_{\mathrm{on}}}{2}\right)\right)-\Theta\left(t-\left(n\tau+\tfrac{t_{\mathrm{on}}}{2}\right)\right)$. Substituting this result into the equation of motion for $\dbtilde{\sigma}_-(t)$ [Eq.~\eqref{eq-supp-mat:sigma-minus-eom-3}] then gives
\begin{equation}
    \tilde{C}(n\tau)=\left[1-\left(gt_\mathrm{on}\right)^2\right]^n\tilde{C}_0(n\tau),\label{eq-supp-mat:ton-decay}
\end{equation}
leading to wavepackets of the form
\begin{equation}
    f_n(t)\simeq -i g t_\mathrm{on}\chi_c(t)=-i g t_\mathrm{on}e^{-i\delta t-\frac{\kappa}{2}t}\Theta(t).
\end{equation}
Following the same reasoning that led to the limits $S_\mathrm{Hahn}$ and $S_\mathrm{CPMG}$ above, we assume the ideal case where $\tilde{C}_0(n\tau)=1$ [in Eq.~\eqref{eq-supp-mat:ton-decay}] and $\left|\braket{\sigma_x}_0\right|=1$. This gives an upper bound
\begin{equation}
    S\le \left[(g t_\mathrm{on})^2 \frac{\kappa_2}{\kappa} N_\mathrm{eff}\right]^{1/2},
\end{equation}
where
\begin{equation}
    N_\mathrm{eff}\le \frac{1}{4}+\sum_{n=1}^\infty \left[1-(g t_\mathrm{on})^2\right]^n\simeq \frac{1}{(g t_\mathrm{on})^2},\quad |g t_\mathrm{on}|\ll 1.
\end{equation}
The signal, being $T_2^*$-independent, is therefore no longer limited by inhomogeneous broadening:
\begin{equation}
    \boxed{S\lesssim S_\mathrm{max} = \sqrt{\frac{\kappa_2}{\kappa}}.}
\end{equation}
In this case, for $\kappa_2/\kappa\to 1$, it is possible (at least in principle) to extract one bit of information per cycle, similar to the case of a single-shot readout.

\end{widetext}
\end{document}